\documentclass[12pt,preprint]{aastex}
\usepackage{graphicx,natbib}

\newcommand{\PKS}{PKS\ 2155-304}
  
\newcommand{\Hnaught}{\ensuremath{H_0}}
\newcommand{\z}{\ensuremath{z}}
\newcommand{\bb}{\ensuremath{b}}
\newcommand{\ang}{~\mbox{\AA}}

\newcommand{\percmtwo}{\ {\rm cm}\ensuremath{^{-2}}~}
\newcommand{\percmtwono}{\ {\rm cm}\ensuremath{^{-2}}}

\newcommand{\perMpc}{~{\rm Mpc}\ensuremath{^{-1}}~}
\newcommand{\perMpcno}{~{\rm Mpc}\ensuremath{^{-1}}}

\newcommand{\nomang}{{\rm m}\mbox{\AA}}
\newcommand{\noang}{\mbox{\AA}}
\newcommand{\mang}{~{\rm m}\mbox{\AA}}

\newcommand{\Ang}{\ang\ }
\newcommand{\Mang}{\mang\ }
\newcommand{\Nno}{\ensuremath{\mathcal{N}}}
\newcommand{\Dperp}{\ensuremath{D_{\perp}}}
\newcommand{\Dtot}{\ensuremath{D_{tot}}\ }
\newcommand{\Dtotno}{\ensuremath{D_{tot}}}

\newcommand{\logNhcm}{\ensuremath{\log{[\Nhno ({\rm cm}\ensuremath{^{-2}})}}]}
\newcommand{\Wno}{\ensuremath{\mathcal{W}}}
\newcommand{\Wnos}{\ensuremath{\mathcal{W}}s}
\newcommand{\W}{\Wno~}
\newcommand{\Ws}{equivalent widths}
\newcommand{\Wi}{\ensuremath{\Wno_i}}

\newcommand{\Lstarno}{\ensuremath{L^*}}
\newcommand{\Lstar}{\Lstarno\ }
\newcommand{\Mstarno}{\ensuremath{M_B^*}}
\newcommand{\Mstar}{\Mstarno\ }

\newcommand{\etno}{et~al.}
\newcommand{\et}{\etno\ }
\newcommand{\eti}{\etno}
\newcommand{\etl}{\et}

\newcommand{\hseventy}{\ensuremath{h_{70}}}
\newcommand{\hsfi}{\ensuremath{h^{-1}_{70}}\ }

\newcommand{\hone}{\ion{H}{1}\ }

\newcommand{\Xtwo}{\ensuremath{\chi^2}\ }

\newcommand{\sig}{\ensuremath{\sigma}\ }
\newcommand{\signo}{\ensuremath{\sigma}}
\newcommand{\sens}{S($\lambda$)\ }
\newcommand{\sensno}{S($\lambda$)}
\newcommand{\onesig}{1\signo\ }
\newcommand{\onesigno}{1\signo}

\newcommand{\threesig}{3\signo\ }
\newcommand{\threesigno}{3\signo}
\newcommand{\foursig}{4\signo\ }

\newcommand{\about}{\ensuremath{\sim}}
\newcommand{\nokmsno}{{\rm km~s}\ensuremath{^{-1}}}
\newcommand{\kmsno}{~\nokmsno}

\newcommand{\sample}{4\signo\Lstar}
\newcommand{\sampleno}{4\signo\Lstarno}

\newcommand{\kms}{\kmsno\ }

\newcommand{\lya}{Ly\ensuremath{\alpha} }
\newcommand{\lyano}{Ly\ensuremath{\alpha}}

\newcommand{\Nhno}{\ensuremath{N_{\rm HI}}}
\newcommand{\Nh}{\Nhno\ }

\newcommand{\definite}{\ensuremath{SL \ge 4\signo}}

\newcommand{\tent}{\ensuremath{3\signo \le SL<4\signo}}
\newcommand{\real}{\ensuremath{SL \ge 4\signo}}

\newcommand{\cDz}{\ensuremath{c\Delta}\z}
\newcommand{\Dz}{\ensuremath{\Delta}\z}

\newcommand{\Dv}{\ensuremath{\Delta v}}
\newcommand{\dz}{\ensuremath{d}\z}
\newcommand{\pz}{\partial\z}

\newcommand{\gt}{\ensuremath{>}}
\newcommand{\lt}{\ensuremath{<}}

\newcommand{\zrange}{\ensuremath{0.002 < \z < 0.069}}
\newcommand{\zem}{\ensuremath{z_{em}}}

\newcommand{\proxno}{c\zem -- 1,200\kmsno}
\newcommand{\highz}{high-\z\ }
\newcommand{\lowz}{low-\z\ }
\newcommand{\lowzno}{low-\z}
\newcommand{\lowzya}{low-\z\ \lya}
\newcommand{\dn}{\ensuremath{d}\Nno}

\newcommand{\pW}{\partial\Wno}
\newcommand{\dndz}{\dn/\dz\ }

\newcommand{\dndzno}{\dn/\dz}

\newcommand{\dndW}{\ensuremath{d\mathcal{N}/d\Wno}}

\newcommand{\dtwondWdzover}{\ensuremath{{\partial^2 \Nno \over \pz~\pW}}}

\newcommand{\cz}{\ensuremath{cz}}

\newcommand{\Exp}[2]{\ensuremath{#1\times10^{#2}}}
\newcommand{\bvalue}{\bb-value\ }
\newcommand{\bvalues}{\bb-values\ }

\newcommand{\SIItriplet}{\ion{S}{2}~\ensuremath{\lambda\lambda}1250, 1253, 1259}

\newcommand{\omegab}{\ensuremath{\Omega_b}}

\newcommand{\Nabs}{81}
\newcommand{\Npos}{30}

\newcommand{\Nsample}{46}
\newcommand{\Nvoid}{8}
\newcommand{\Nsuper}{38}

\newcommand{\Msolar}{\ensuremath{M_{\odot}}}
\newcommand{\vheliono}{\ensuremath{V_{\rm hel}}}
\newcommand{\vhelio}{\vheliono\ }
\newcommand{\Voidfraction}{$22 \pm 8$\%}
\newcommand{\baryonfraction}{$4.5 \pm 1.5$\%}

\begin{document}

\title{The Local \lya Forest. III. Relationship between \lya Absorbers \\
   and Galaxies, Voids and Superclusters 
\footnote{Based on observations with the NASA/ESA Hubble Space Telescope, 
   obtained at the Space Telescope Science Institute, which is operated 
   by the Association of Universities for Research in Astronomy, Inc. 
   under NASA contract No. NAS5-26555.}  }

\author{Steven V. Penton, John T. Stocke, and J. Michael Shull 
\footnote{Also at JILA, University of Colorado
   and National Institute of Standards and Technology.}}
\affil{Center for Astrophysics and Space Astronomy, 
   Department of Astrophysical and Planetary Sciences,
   University of Colorado, Boulder CO, 80309}

\email{spenton@casa.colorado.edu, stocke@casa.colorado.edu, and
   mshull@casa.colorado.edu}

\shorttitle{HST/GHRS observations of the \lowya Forest}

\shortauthors{Penton, Stocke, \& Shull}

\begin{abstract}

In this paper, we use large-angle, nearby galaxy 
redshift surveys to investigate the relationship 
between the \Nabs\ low-redshift \lya absorbers in our HST/GHRS
survey and galaxies, superclusters, and  voids.  
In a subsample of \Nsample\ \lya absorbers located in
regions where the February 8, 2000 CfA catalog is
complete down to at least \Lstar galaxies, the nearest
galaxy neighbors range from $100\hsfi $kpc to $\gt 10\hsfi $Mpc.
Of these \Nsample\ absorbers, \Nvoid\ are found in galaxy voids.
After correcting for pathlength and sensitivity, we find that \Voidfraction\
of the \lya absorbers lie in voids, which
requires that at least some low-column density absorbers are
not extended halos of individual bright galaxies. The number density of
these clouds yields a baryon fraction of \baryonfraction\ in voids.

The stronger \lya absorbers ($10^{13.2-15.4}\percmtwono$) cluster with galaxies more weakly
than galaxies cluster with each other, while the weaker absorbers ($10^{12.4-13.2}\percmtwono$)
are more randomly distributed.  
The median distance from a \lowz \lya absorber in our sample to its nearest
galaxy neighbor ($\about 500 \hsfi $kpc) is twice the median distance between
bright galaxies in the same survey volume. This makes any purposed
``association'' between these \lya absorbers and individual galaxies problematic.
The suggested correlation between \lya absorber equivalent width (\Wno) and nearest-galaxy impact parameter
does not extend to $\Wno \le$ 200\mang, or to impact parameters
$\gt200\hsfi $kpc. Instead, we find statistical support for the
contention that absorbers align with large-scale filaments of
galaxies. The pair of sightlines, 3C~273 and Q~1230+0115, separated by
0.9\degr\  on the sky, provides an  example of
8 absorbers and 7 galaxies aligned along a possible filamentary structure at least
20\hsfi Mpc long. While some strong ($\Wno \gt 400$\mang) \lya absorbers may 
be gas in the extended gaseous halos of
individual galaxies, much of the local \lya ``forest'' appears to be
associated with the large-scale structures of
galaxies and some with voids.  

\end{abstract}

\section{Introduction}\label{sec:obs}
Since the discovery of the high-redshift \lya forest over 25 years
ago, these abundant absorption features in the spectra of QSOs
have been used as evolutionary probes of the intergalactic medium
(IGM), galaxy halos, large-scale structure, and chemical evolution. 
In the past few years, these discrete \lya lines have been interpreted
theoretically by N-body hydrodynamical models \citep{Cen94,Hernquist96,Zhang97,Dave99}
as arising from baryon density fluctuations associated with gravitational
instability during the epoch of structure formation.  

Most previous studies of the \lowz IGM \citep{Bahcall93,Jannuzi98,Weymann98}
have been performed with the Faint Object Spectrograph (FOS) aboard the Hubble Space
Telescope (HST).  These studies, in general, characterized the \lya absorbers
with rest-frame equivalent widths (\Wno) greater than 0.24\ang.  
These studies detected some features with $\Wno\le240\mang$, but,
because of sensitivity function considerations,
performed most of their \lya statistics for $\Wno\ge 240\mang$.
A great deal more information can be gained from 
studying the more numerous weak \lya lines,  
using the moderate-resolution first-order gratings (19\kmsno) on the HST spectrographs,
either the G160M grating on the Goddard 
High Resolution Spectrograph (GHRS) or the Space Telescope Imaging 
Spectrograph (STIS) with medium resolution first-order gratings.  
With these instruments,
the Colorado group has engaged in a long-term program to study
the very low redshift ($z < 0.069$) \lya forest. 
In Paper~I \citep{PaperI} we described our
GHRS/G160M observations and presented a catalog of \lya absorbers 
toward 15 extragalactic targets.
We detected
\Nabs\ ``definite'' \lya absorbers at a significance 
level $\geq4\sigma$
and \Npos\ additional ``possible'' absorbers at $3-4\sigma$ 
in the redshift range \zrange\ over a total pathlength
 \cDz = 116,000\kms (\Dz = 0.387).  
In Paper~II \citep{PaperII} we described the 
physical properties of these \lya absorbers and compared them
to their high-redshift counterparts. 

An important result from Paper~II  is that the number density of absorbers per unit 
redshift rises sharply at \lya rest-frame equivalent widths $\Wno \le 100\Mang$ 
($\Nh \le 10^{13.4}\percmtwo$ for Doppler parameter $\bb = 25\kmsno$; see Figure~8 in Paper~II).
This corresponds to a region in column density probed only by observations made with the 
GHRS or STIS and medium resolution first-order or echelle gratings. Both 
the column density distribution and its evolution with redshift
suggest that these lowest column density clouds are a population physically 
distinct from those at higher column density, as has been widely discussed for the \highz
\lya forest for some time \citep[e.g.,][]{Sargent87}. The two-point correlation function (TPCF) of local \lya absorbers has no excess power over a
random distribution except for $\Delta v \leq 200\kmsno$, where a modest excess of close pairs
 is observed (Figure~21 in Paper~II, and
Impey, Petry \& Flint 1999; hereafter IPF99). As we show in \S~\ref{sec:GGbeam}, the \lya TPCF
has a much lower amplitude than the galaxy-galaxy TPCF at \z=0.

Several previous investigations used the
detection of relatively nearby \lya absorbers to determine
whether a physical relationship exists between \lya absorbers
and galaxies. Using HST Key
Project data \citep{Bahcall93,Jannuzi98}, 
\citet[L95 hereafter]{Lanzetta95} proposed that
most, if not all, \lowzya absorbers arise in extended
halos of galaxies. This claim, limited to those absorbers
detected in FOS data ($\Wno \ge 0.24$\ang), was based
upon an observed inverse correlation between \W and the
impact parameter from the QSO sightline to the nearest galaxy. 
An extension of the L95 work by \citet[C98 hereafter]{Chen98} strengthened
the observed inverse correlation but found little evidence
that this correlation depended on any other galaxy property, aside from galaxy luminosity.
This is  contrary to what would be expected if the size of the gaseous 
galaxy halo were due to internal properties or structures in the galaxies (e.g., star
formation rate and thus galaxy color, Hubble type, etc.). 
In the most recent work by this group \citep{Chen01}, the authors conclude that all
galaxies, regardless of Hubble type or recent star formation history, possess extensive gaseous
halos of near unity covering factor; e.g., an \Lstar galaxy has a halo extent of ~260\hsfi
kpc. We assume a Hubble constant of \Hnaught=70 \hseventy\kmsno\perMpc
throughout the paper.
The limitations of the L95 and C98 work
are that their sample was based on only the strongest absorbers at low \z,
and that they did not consider absorbers for which no nearby galaxy was detected.
Based upon a detection of a nearby
galaxy in only about 1/3 of the cases, L95 and C98 extrapolate
their result to all nearby \lya absorbers, assuming that
absorbers for which no nearby galaxy was detected are associated
with galaxies too faint to be discovered using their survey.

\citet{Tripp98a} and \citet{Stocke95} claim that neither of the C98 extrapolations 
are justified and that the absence of correlations with other
galaxy properties argues against the ``galaxy halo" model.
Numerical simulations  \citep{Dave99}
are able to reproduce the L95 inverse correlation without 
extended galaxy halos, because it arises  from the presence of both galaxies and \lya
absorbers within the same large-scale filaments.
Therefore the connection suggested by L95 may be only
coincidental, not physical or causal. The
simulations also suggest that weaker absorption lines
will be found farther from galaxies and will fail to show the inverse
correlation seen for the stronger absorbers, as has
been observed (Stocke \etl 1995; Tripp, Lu, \& Savage 1998a; IPF99).

By using \lya lines detected with the GHRS, \citet{Morris93}, 
\citet{Stocke95}, \citet{Tripp98a}, \citet{Grogin98}, and IPF99  investigated
the relationship between weak \lya absorbers and
galaxies. The better far-UV sensitivity of the GHRS allowed these
studies to detect absorbers at substantially lower redshifts
and thus much nearer than typical Key Project detections
(i.e., at \z\ $\leq 0.2$ compared to $0.1\leq z \leq 1.6$).
The superior GHRS resolution allowed the detection of much weaker
absorbers ($\Wno\geq0.1$\Ang for the low-resolution
grating studies of \citet{Tripp98a} and IFP99 
 and $\Wno \geq 0.015$\Ang for the medium-resolution
studies of \citet{Morris93} and this work) albeit over shorter
pathlengths than the Key Project data. 

None of these studies found any compelling evidence for a close association between
galaxies and absorbing clouds. On the contrary, the initial study
of \citet{Morris93} along the sightline to 3C~273, found
nearest bright-galaxy neighbors at $\geq$ 200 \hsfi kpc, with
some nearest galaxies more than 1\hsfi Mpc away from some clouds.
Followup scrutiny of the nearest of these clouds, in the southern
extremities of the Virgo Cluster at \cz=1000--1600\kmsno,
failed to find any galaxies within 100-150 \hsfi kpc of these
absorbers down to impressively low limits ($M_B\geq-13$; Rauch,
Weymann \& Morris 1996).
These include low surface brightness
galaxies, as would be detected optically at central surface
brightness limits of $\geq 26.4$ mag arcsec$^{-2}$ \citep{Rauch96}
or by 21~cm emission 
\citep[$M_{HI}\geq10^{7}~h_{70}^{-2}~M_{\odot}$;][]{VG93}. 
Substantiating these results, \citet{Stocke95} and \citet{Shull96}
found some nearby, \lowzno, \lya absorbers in ``voids" in the bright CfA
galaxy redshift survey regions, with nearest bright galaxies several Mpc
distant. Because both the brighter and the fainter galaxies ``avoid
the voids", these authors concluded that at
least some local \lya absorbers were not associated with galaxies
at all. A nearest-neighbor galaxy analysis using the bright
survey data suggested that, while these low column density
absorbers do not cluster with galaxies as strongly as galaxies cluster
with other galaxies,
neither are they randomly distributed. They are slightly
closer to galaxies than would be expected by chance. This result
confirmed the earlier statistical results of \citet{Morris93},
 using the 3C~273 sightline 
\citep[see][for an analysis of this result in terms of several
populations of clouds, some closely associated with galaxies, 
some not at all associated]{Mo94}. \citet{Stocke95} suggested that
these results are compatible with \about~2/3 of these absorbers
being associated with large-scale structures (``filaments")
in the galaxy distribution with the remaining absorbers distributed entirely
randomly. The much lower amplitude in the TPCF of these 
low-\Nh absorbers, compared to the galaxy TPCF,
 is consistent with the filament interpretation (IPF99, Paper~II).

More recently, \citet[GG98 hereafter]{Grogin98} used a
small number (18) of \lya absorbers found in the region of the
CfA galaxy redshift survey to determine that \lya absorbers are
found at locations random with respect to galaxy density. 
Their smoothing kernel of 2--5~\hsfi Mpc allowed them
to reconstruct a robust measure of Galactocentric galaxy density, compared to
using the nearest-neighbor galaxy distance, which is prone to biases due
to the magnitude limits of the galaxy survey employed.
However, these smoothing lengths may be too large to assess accurately the possible
physical association between absorbers and large-scale structure
suggested by earlier work. 
IPF99 addressed this difficulty by determining both
nearest-neighbor distances and smoothed galaxy densities for 11
\lya absorbers in the Virgo supercluster region using a galaxy
survey complete to $M_B\leq-16$. This impressive work finds that,
while there is some bias in using nearest-neighbor distances to
\Lstar galaxies, the distances to fainter galaxies ($L\geq0.04$\Lstarno)
are still too large (240-1320 \hsfi kpc for all but one absorber) to be
considered as galaxy halos. Further, when smoothed using
2~\hsfi Mpc spheres, \lya absorbers are found in intermediate, not
random, regions of galaxy density, supporting earlier
speculations to this effect by \citet{Mo94} and \citet{Stocke95}. 
While IFP99 is the most comprehensive study of the
relationship between \lya absorbers and galaxies to date, it is
still limited by the small number of absorbers studied (11) and the
location of these absorbers within the Virgo supercluster region
alone.

In this paper (Paper~III), we use the \Nabs\ \lya absorbers, whose
discovery was reported in Paper~I and analyzed in Paper~II, in
conjunction with currently available galaxy redshift survey data to
explore the relationship between  the \lowzya absorbers and bright
galaxies.  These sightlines are scattered across the sky, with only two
targets behind the Virgo supercluster (3C~273 and Q~1230+0115). 
Our analysis uses  the  nearest-neighbor galaxy distribution of our
absorber sample and also looks for correlations between the \lya
absorbers and large-scale galaxy structures (supercluster filaments
and voids) that these galaxy surveys reveal. In \S~\ref{sec:SAG}, we
describe our
\lya sample and the galaxy catalog, which is based on  the February 8,
2000 version of the Center for Astrophysics  (CfA) redshift survey, plus
a few other ``pencil-beam'' optical and \hone  surveys  along specific
sightlines using multi-object optical spectroscopy or 21~cm imaging
spectroscopy  from the Westerbork Synthesis Radio Telescope  (WSRT) and
the Very Large Array (VLA).  

To examine the relationship of the \lya absorbers to the 3-D
galaxy distribution  along each sightline, we use the ``pie diagrams'' in
right ascension and declination and the tables of nearest galaxy neighbors
presented in Paper~I.  We also develop a galaxy-absorber sample 
that is consistent in galaxy coverage, to avoid biases due
to galaxy survey incompleteness. Our restricted sample 
(called the \sample sample) is composed 
of \Nsample~\lya absorbers of $\geq 4\sigma$ significance, located in
regions of the local Universe where the galaxy catalogs are
complete down to \Lstar galaxies, or below.
In \S~\ref{sec:assoc}, we use this sample to examine galaxy--absorber 
``coincidences'' and the distribution of nearest-neighbor galaxies.  
We combine our galaxy detections and non-detections with the 
results of several other surveys in an attempt to determine 
whether most \lya absorptions can be explained as arising from 
extended galaxy halos.  We also compare the nearest-neighbor 
distribution of our \sample sample to the galaxy-galaxy 
nearest-neighbor distribution, and to a random-galaxy distribution
within the same galaxy survey regions.  
In \S~\ref{sec:filaments} we explore the relationship between \lya absorbers and galaxy 
filamentary structures, both in terms of an individual example, the
close pair of sightlines towards 3C~273 and Q~1230+0115, and by using
statistical methods on the GHRS \sample sample. 
In \S~\ref{sec:dndz_voids} we explore the properties of \lya absorbers in galaxy
superclusters and  voids in terms of the \lya absorber frequency, $d \mathcal{N}/dz$, as a function of
\Wno. In \S~\ref{sec:GGbeam} we compare the two-point correlation function of galaxy halos 
to that  derived in Paper~II for \lya absorbers.
Finally, \S~\ref{sec:conclusions} summarizes the most important conclusions
 drawn from these various comparisons and outlines work in
progress that will improve the current results (e.g., 
in Penton, Stocke \& Shull, 2002, we will
incorporate an additional 15 sightlines observed with HST/STIS).
\clearpage
\section{Sample of Absorbers and Galaxies} \label{sec:SAG}
\subsection{CfA Galaxy Pie Charts} \label{sec:pie}
In Paper~I, we presented heliocentric ``pie diagrams'' 
in right ascension and declination
versus heliocentric radial velocity for all 15 HST 
targets in our GHRS survey. These pie diagrams indicate the 
spatial positions of the CfA Redshift Catalog galaxies relative 
to our target sightlines and the detected \lya
absorbers, assuming a pure Hubble flow. The February 8, 2000 version of 
this catalog \citep{CFA} contains \about120,000 galaxies with 
velocities less than 100,000\kmsno, including several large sky areas where complete,
magnitude-limited surveys have been completed (see \S~\ref{sec:sample}). 

All \lya absorber velocities were reported in Paper~I 
in the LSR velocity scale, established by
aligning the velocity centroids of the Galactic \SIItriplet~absorption 
lines with the dominant Galactic 21-cm emission. 
We selected this velocity scale instead of the HST/GHRS wavelength 
solution owing to the possibility of an improper 
wavelength scale caused by poor target centering in the large science 
aperture (LSA). 
To convert our LSR \lya absorber velocities to heliocentric
velocities, we assumed that the solar velocity with respect to LSR 
is +20.0\kms towards  ($\alpha$=18:03:50.3;$\delta$=+30:00:17; epoch J2000). 
Due to this process, we estimate
conservatively that the typical velocity accuracy of the \lya absorbers is $\pm10\kmsno$. Our accuracy is limited by
the  centroid uncertainty of the Gaussian fit, added in 
quadrature with the coincidence accuracy of the dominant \hone 21-cm emission and the
S~II absorption lines. The error in the mean
of the Gaussian fit is listed individually for each line detected in
Table~2 of Paper~I. The accuracy of the CfA redshift catalog galaxy recession velocities is
variable (median velocity error of 34\kmsno; mean velocity error of 43\kmsno).
Herein, the absorbers of Paper~I are converted to heliocentric velocities to match the galaxy data.

\subsection{Nearest-Neighbor Galaxies} \label{sec:NN}
\begin{figure}[htb]
  \epsscale{0.8}
  \plotone{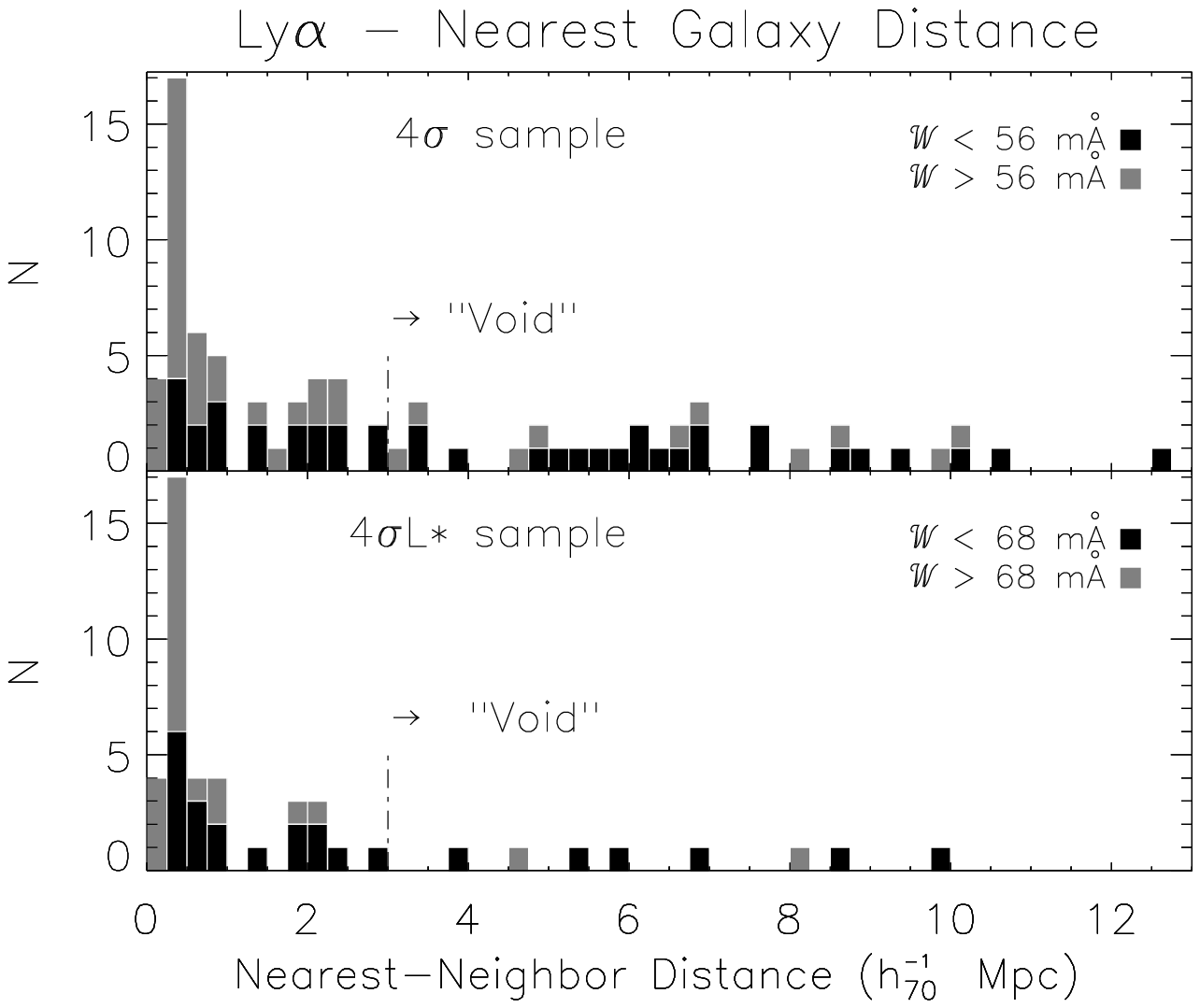}
\caption{\label{NNhist} The upper panel displays a histogram of the nearest-neighbor distance to a CfA catalog galaxy 
for our \Nabs\ \definite\ \lya absorbers, split at the median \W (56\mang).  
Each 250\hsfi kpc bin should be considered an upper limit on 
3D nearest-neighbor distance (\Dtotno), since the location of all 
galaxies is not known. Because our sightlines pass through regions 
of varying completeness in the CfA catalog, this distribution is 
non-uniform and possibly skewed towards larger nearest-neighbor 
distances. The lower histogram is for our restricted \sample \lya sample (46 absorbers, see
\S~\ref{sec:sample}), split at its median \W (68\mang). In both median split samples, 
a K-S test indicates that the two subsamples are drawn from 
different parent populations at $\gt$ 99.3\% significance level. 
We identify the 8 absorbers in the lower (\sampleno) panel  with nearest-neighbor 
distances \gt 3 \hsfi Mpc as ``void absorbers''.}
\end{figure}

In the upper panel of Figure~\ref{NNhist}, we display a histogram of the nearest-neighbor 
distances from each of our \Nabs\ \foursig absorbers to the nearest CfA redshift survey
galaxy. The distribution is split at the median \W (56\mang, or $\about10^{13.1}$\percmtwo for $\bb=25$\kmsno).
The dark bins are the histogram for the weaker absorbers,
while the grey bins are the stronger absorbers. 
A two-sided  Kolmogorov-Smirnov (K-S) test indicates that the weaker and stronger
distributions are different at the 99.9\% ($4\sigma$) confidence level. 
We suspect that the weaker and stronger absorbers may sample different populations 
with different space distributions.

In Paper~I we listed the nearest 3 galaxies for all SL\gt\threesig \lya 
absorbers in our sample, with information on galaxy morphology, location, dimensions,
velocity,  and magnitude taken from the Revised CfA Redshift
Catalog unless otherwise indicated. The line of sight (LOS) distance from
the absorber to the galaxy uses a ``retarded Hubble 
flow'' model \citep{Morris93,Stocke95} which accounts for peculiar galaxy motion and
galaxy rotation by decreasing the reported total absorber-galaxy 
distance compared to a ``pure Hubble flow'' model where LOS distance is
given by $c(z_{abs}-z_{gal})\Hnaught^{-1}$. In the ``retarded Hubble flow" model,
we consider any galaxy within  $\pm$~300\kms of the recession velocity 
of an absorber to be at the same distance from us as the absorber, 
and we decrease all LOS distances by $c(\z_{abs}-\z_{gal})\Hnaught^{-1}$, 
if $c(\z_{abs}-\z_{gal})<300\kmsno$, or by 300\kms if $c(\z_{abs}-\z_{gal}) > 300\kmsno$.
All three-dimensional distances quoted herein use this ``retarded Hubble
flow" model for computing distances long the line-of-sight and assume
Euclidean geometry, given the small redshifts involved. Therefore, our
3D nearest-neighbor distances should be considered conservative lower
limits.

While the $\pm 300\kms$ tolerance on matching absorber to galaxy velocities is
a somewhat arbitrary choice, we note several reasons that brought us to this selection.
First, halo stars (L95) and high velocity gas associated with our own galaxy extend to
these velocities. Secondly, the rotation curves of the most massive galaxies extend to
these velocities. While this value is somewhat greater than typical velocity
dispersions of galaxy groups and supercluster ``filaments'', it is typical of the full
range of velocities therein \citep[e.g.,][]{Ramella92}. 
Finally, while L95 allowed a $\pm250\kms$ leeway for ``association'' between absorbers and galaxies, both
the FOS absorber velocities and the faint galaxy velocities obtained by L95 are less
accurate than those used here. So, to be conservative in the sense of being more
inclusive of possible absorber-galaxy ``associations'' (and thus more open to the L95
hypothesis that all absorbers are extended galaxy halos), we use this somewhat 
larger range for absorber-galaxy velocity differences. In a ``pure Hubble flow'' model
(i.e., no velocity tolerance), the 3D distances between galaxy and absorber would be even larger.
\subsection{The \sample Sample} \label{sec:sample}
Because the CfA redshift survey is a flux-limited survey and does 
not contain uniform galaxy information as a function of position on the
sky,
some of our sightlines are better suited than others for nearest-neighbor 
analysis.  Four of our 15 sightlines (I~ZW~1, MRK~335, 
MRK~421, and MRK~501) 
pass through well-surveyed regions of the February 8, 2000 CfA survey. In an earlier publication
\citep{Stocke95}  we used these sightlines to investigate probable galaxy-cloud relationships.
To date, the well-surveyed regions of the CfA catalog 
that are complete to $m_B  \leq 15.5$ are (in 1950 coordinates):
$$8^h \ \ \leq \alpha \ \leq \ \ 17^h~~;~~2.\degr5 \ \leq \ \delta \ 
   \leq \ 44.\degr5 \qquad ~~{\rm First~6~CfA~Slices}$$
$$20^h \ \ \leq \alpha \ \leq \ \ 4^h~~;~~-2.\degr 5 \ \leq \ \delta \ 
   \leq \ 90\degr \qquad {\rm Southern~Galactic~Cap}$$
$$4^h \ \ \leq \alpha \ \leq \ \ 8^h~~{\rm and}~~17^h \ \ \leq \alpha \ 
   \leq \ \ 20^h~~;~~0\degr \leq \ \delta \ \leq \ 45\degr \qquad {\rm [LGL].}$$
Galaxies in the low Galactic latitude survey (LGL) are taken from 
\citet{Marzke96}, while the Southern Galactic Cap data are 
taken from \citet{Huchra99}. 
The CfA catalog also includes all galaxies in the Morphological Catalog 
of Galaxies (MCG, Huchra \etl 1993),  complete to $m_B~\leq~15.0$ in the 
following region:
$$ 20^h \ \ \lt \alpha \ \lt \ \ 5^h~~;~~-17.\degr 5  \ \leq \ \delta \ 
   \leq \ -2.\degr 5 \qquad {\rm [MCG].} $$
One sightline (MRK~509) lies in this region of the sky.  
The CfA catalog also includes all galaxies in the merged Zwicky-Nilson 
catalog (Nilson 1973),  complete to $m_B~\leq~14.5 $ in the regions:
$$b \ \geq \  40\degr ~~;~~\delta \ \geq \  0\degr ~~ {\rm and }~~b \ 
   \leq \-30\degr ~~;~~\ \ \delta \ \geq \  -2.\degr 5\ \qquad {\rm [NZ40]}.$$
This region is known as the North Zwicky 40 (NZ40), and it includes the sightlines 
of MRK~279, MRK~290, and MRK~817. In addition, 
\citet{GGH98} (GGH) have extended the CfA survey down to $m_B$=15.7 in the region:
$$ 11.5^h \ \ \leq \alpha \ \leq \ \ 13.5^h~~;~~-3.\degr 5 \ \leq \ 
   \delta \ \leq \ 8.\degr 5 \qquad {\rm [GGH],}$$
which includes the sightlines towards 3C~273 and Q~1230+0115. Finally, 
the second southern sky redshift survey (SSRS2; da Costa 1999) contains
the \PKS\ sightline and is complete down to $m_B~\leq~15.5 $ in the region:
$$\left(-40\degr \leq  \delta\ \le\ 2.\degr 5~~;~~b \leq \ 
   -40\degr\right) ~~{\rm or }~~ \left( \delta \le 0\degr ~~;~~ 
   b \ge 35\degr\right) \qquad {\rm [SSRS2].}$$
The CfA catalog also includes the Bright Galaxy Redshift Catalogue (BGRC), which is an 
all-sky catalog complete to $m_B$=13.21.

In addition, five of our objects (3C~273, H~1821+643, MRK~335, MRK~501, 
and \PKS) have undergone optical multi-object spectroscopy and/or
``pencil-beam'' \hone surveys with WSRT or VLA.
The 3C~273 sightline was observed by \citet{Morris93} with the Fiber 
Spectrograph on the Las Campanas 2.5m du Pont Telescope using the 
2D-FRUTTI detector. Redshifts were obtained over 
region of radius $>1\degr$ centered on 3C~273. The H~1821+643 sightline was observed 
by \citet{Tripp98a} using the WIYN+HYDRA in a 1\degr\  circular field of 
view. These optical surveys are reported to be complete down to 
$m_B \leq 19$ over a full 1\degr\ field and 72\% complete to $m_B \sim 18$ over a 20' radius,
respectively.  However, as with any fiber survey, some galaxies might be 
missed due to fiber positional misalignments and positioning constraints. 
The MRK~335 and \PKS\ sightlines were observed with the VLA 
(van Gorkom \etl 1996; Shull \etl 1998). The \PKS\ 
sightline was observed over the velocity ranges 3562--6637\kms and 
16,230--17,530\kms to an \hone mass limit well below that of 
an \Lstar spiral galaxy.  MRK~335 was observed with the VLA over 
the velocity range 1635--2635\kms at a resolution of 25\kmsno. The 
MRK~501 sightline was observed over the velocity ranges 
4385--5385\kms and 7240--8240\kms with a resolution of 17\kms \citep{VG96}.
Table~\ref{completeness} summarizes the galaxy information available 
for each sightline, including by column: (1) sightline target name; (2) the Zwicky (blue)
magnitude limit of the catalog along this sightline; (3) the catalog names as described
above; and (4) the velocity limit ($cz_c$) at which the catalog magnitude limit equals \Lstarno.
Velocities along the sightline less than the velocity limit ($cz_c$) are complete to magnitudes
fainter than \Lstarno.

Owing to our concerns over \lya absorbers with significance levels 
\tent, we only include absorbers with \real\ in our analysis.
To achieve some consistency in our nearest-neighbor sample, we adopt the
following condition for inclusion in our \sample sample: the absorber must 
reside at a velocity and location on the sky in which the CfA redshift 
catalog (or other survey)
is inclusive of all galaxies down to at least \Lstarno. 
In the Zwicky (blue) magnitude system, an \Lstar
galaxy has an absolute magnitude of \Mstar = 
$5 \log \hseventy - 19.57$ \citep{Marzke94}; 
for the objects in the well-surveyed CfA regions, which are complete 
for $m_B \leq 15.5$, this velocity cutoff is $\cz \le 7230$\kmsno. 
Table~\ref{completeness} summarizes the velocity restrictions for each of 
our sightlines. Note that the \PKS\ sightline
is unusual, since its velocity coverage is not contiguous.
Under these velocity and significance level restrictions, there are 
\Nsample~absorbers in our \sample sample.
Table~\ref{linelist_sorted_def} presents all \foursig absorbers, 
sorted by velocity with the following information by column: 
(1) target name (an asterisk indicates pre-COSTAR data), 
(2) the absorber heliocentric radial velocity and its error in\kmsno, 
(3) the rest-frame equivalent width (\Wno) and its error in \mang, 
(4) whether this absorber is included in the \sample sample, 
(5) the CfA redshift catalog name of the nearest galaxy to this absorber, 
(8) the total Euclidean absorber-galaxy distance (\Dtotno, assuming the ``retarded Hubble flow'' model;
 see \S~\ref{sec:sample}),
(6) the apparent Zwicky blue magnitude of this galaxy, and
(7) the heliocentric radial velocity of this galaxy.
When calculating absorber-nearest galaxy distances, we do not restrict ourselves to \Lstar
or brighter galaxies, but consider all known galaxies in the field. Notice that, owing to
the peculiar velocity allowance ($\pm 300\kmsno$) assumed here, several
absorbers are paired with the same nearest known galaxy.
	\setlength{\tabcolsep}{2mm}
	\tableheadfrac{0.05}
	\renewcommand{\arraystretch}{0.75}
	\begin{deluxetable}{lcrc}
\tablecaption{Completeness Limits for our \sample
Sample.\label{completeness}}
\tabletypesize{\scriptsize}
\tablecolumns{4}
\tablewidth{0pt}
\tablehead{
\colhead{Target} &
\colhead{m$_b$\tablenotemark{a}} &
\colhead{Catalog\tablenotemark{b}} &
\colhead{$cz_c$ Limit\tablenotemark{c}~ (\kmsno)} 
}
\startdata
3C273&19.00&GGH \& \citet{Morris93}\tablenotemark{d}&$cz_c \lt 36233$\\
AKN120&13.21&BGRC&$cz_c$ $\lt$ 2519\\
ESO141-G55&13.21&BGRC&$cz_c$ $\lt$ 2519\\
FAIRALL9&13.21&BGRC&$cz_c$ $\lt$ 2519\\
H1821+643&18.00&\citet{Tripp98a}\tablenotemark{e}&$cz_c \lt 22784$\\
IZW1&15.50&South Gal Cap&$cz_c \lt 7230$\\
MARK279&14.50&N Zwicky 40&$cz_c \lt 4562$\\
MARK290&14.50&N Zwicky 40&$cz_c \lt 4562$\\
MARK335&15.50&South Gal Cap&$cz_c \lt 7230$\\
MARK335&\nodata&VLA - \citet{VG96} &$1635 \lt cz_c \lt 2635 $\\
MARK421&15.50&2nd CfA Slice&$cz_c \lt 7230$\\
MARK501&15.50&1st CfA Slice&$cz_c \lt 7230$\\
MARK501&\nodata&WSRT; \citet{VG96}&$cz_c \lt 8240$\\
MARK509&15.00&MCG&$cz_c \lt  5743$\\
MARK817&14.50&N Zwicky 40&$cz_c \lt 4562$\\
PKS2155-304&15.50&SSRS2&$cz_c \lt  7230$\\
PKS2155-304&\nodata&VLA $HI$&$16230 \lt cz_c \lt 17571$\\
Q1230+0115&15.70&GGH&$cz_c \lt 7927$\enddata
\tablenotetext{a}{Blue (Zwicky) magnitude limit of the indicated catalog along this sightline.}
\tablenotetext{b}{Galaxy catalog used in determining nearest neighbors. The Bright Galaxy Redshift Catalog
(BGRC), Southern Galactic Cap, North Zwicky 40, GGH, SSRS2, MCG, and CfA slices are taken from the February 8,
2000 version of the CfA redshift catalog; see text for acronym definitions.}
\tablenotetext{c}{Velocity cutoff limit for which the indicated catalogs are complete down to \Lstar galaxies,
assuming \Hnaught = 70\kmsno\perMpcno.}
\tablenotetext{d}{CfA + 1\degr-radius 2D-FRUTTI galaxy survey complete to $m_B \about 19$.}
\tablenotetext{e}{CfA + 20'-radius WIYN/HYDRA galaxy survey 72.4\% complete at $m_B < 18$.}
\end{deluxetable}

\clearpage
	\begin{deluxetable}{lccccccc}
\tablecaption{Definite (\real) \lya  features, listed in order of increasing velocity.\label{linelist_sorted_def}}
\tabletypesize{\scriptsize}
\tablewidth{0pt}
\tablecolumns{8}
\tablehead{
\colhead{Target} &
\colhead{$cz_{abs}$} &
\colhead{\W\tablenotemark{a}}&
\colhead{In\tablenotemark{b}~~\sample} &
\colhead{Nearest} &
\colhead{D$_{tot}$\tablenotemark{c}}&
\colhead{m$_B$\tablenotemark{d}} & 
\colhead{$cz_{gal}$} \\
\colhead{} &
\colhead{(\nokmsno)} &
\colhead{(m\AA)} &
\colhead{ sample ?} &
\colhead{Galaxy}&
\colhead{(Mpc)}&
\colhead{}&
\colhead{(\nokmsno)}
}
\startdata
$*$3C273&  1013 $\pm$     6& 369 $\pm$  36&Y&12285+0157 & 0.17&15.60& 1105 \\ 
Q1230+0115&  1487 $\pm$     6& 138 $\pm$  42&Y&A1225+0152 & 0.27&16.30& 1298 \\ 
$*$3C273&  1583 $\pm$     6& 373 $\pm$  30&Y&A1225+0152 & 0.18&16.31& 1298 \\ 
Q1230+0115&  1664 $\pm$     9& 385 $\pm$  94&Y&12308+0147 & 0.28&18.55& 1643 \\ 
Q1230+0115&  1743 $\pm$     7& 241 $\pm$  99&Y&12308+0147 & 0.28&18.08& 1643 \\ 
Q1230+0115&  1857 $\pm$    13& 142 $\pm$  81&Y&12308+0147 & 0.28&17.31& 1643 \\ 
MARK817&  1919 $\pm$     9&  29 $\pm$  13&Y&14332+5934 & 0.28&18.45& 1920 \\ 
$*$MARK335&  1961 $\pm$     6& 229 $\pm$  30&Y&00025+1956 & 0.10&18.45& 1950 \\ 
MARK817&  2083 $\pm$     5& 135 $\pm$  15&Y&14332+5934 & 0.28&18.45& 1920 \\ 
$*$3C273&  2287 $\pm$     7&  35 $\pm$  30&Y&A1225+0258 & 0.40&17.10& 2329 \\ 
$*$MARK335&  2291 $\pm$    12&  81 $\pm$  26&Y&N7817      & 0.45&18.88& 2310 \\ 
Q1230+0115&  2298 $\pm$     6& 439 $\pm$  57&Y&A1228+0116 & 0.16&18.88& 2289 \\ 
MARK509&  2549 $\pm$     6& 209 $\pm$  32&Y&A2028-1821 & 4.66&17.99& 2314 \\ 
$*$PKS2155-304&  2630 $\pm$    15&  42 $\pm$  40&Y&A2156-3128 & 0.70&\dots& 2786 \\ 
$*$PKS2155-304&  2782 $\pm$    16&  36 $\pm$  22&Y&A2156-3128 & 0.70&15.60& 2786 \\ 
MARK421&  3035 $\pm$     6&  86 $\pm$  15&Y&A1102+4120 & 2.14&\dots& 2988 \\ 
$*$PKS2155-304&  4028 $\pm$    12&  21 $\pm$  11&Y&N7201      & 2.77&\dots& 4452 \\ 
AKN120&  4057 $\pm$     8&  48 $\pm$  18&N&A0510-0036 & 0.87&\dots& 4284 \\ 
$*$MARK335&  4265 $\pm$    14&  33 $\pm$  16&Y&00008+2150 & 2.25&16.10& 4455 \\ 
MARK290&  4652 $\pm$     7&  60 $\pm$  18&N&A1542+5722 & 1.65&\dots& 4287 \\ 
$*$MARK501&  4659 $\pm$    10& 161 $\pm$  43&Y&16510+3927 & 0.52&14.42& 4625 \\ 
MARK817&  4668 $\pm$    10&  23 $\pm$  11&N&14293+5528 & 6.54&\dots& 5291 \\ 
$*$PKS2155-304&  4949 $\pm$     7&  64 $\pm$  23&Y&A2157-3025 & 0.33&16.30& 5187 \\ 
$*$PKS2155-304&  5010 $\pm$     7&  82 $\pm$  22&Y&A2157-3025 & 0.33&16.30& 5187 \\ 
MARK817&  5074 $\pm$     6& 207 $\pm$  14&N&14293+5528 & 4.78&16.30& 5291 \\ 
$*$PKS2155-304&  5116 $\pm$     7& 218 $\pm$  20&Y&A2157-3025 & 0.33&16.30& 5187 \\ 
MARK817&  5222 $\pm$     7&  25 $\pm$   7&N&14293+5528 & 4.78&17.40& 5291 \\ 
MARK279&  5234 $\pm$     7&  30 $\pm$  10&N&A1310+6745 & 8.89&17.40& 5994 \\ 
$*$PKS2155-304&  5615 $\pm$     7&  29 $\pm$  15&Y&A2157-3025 & 1.86&15.70& 5187 \\ 
$*$PKS2155-304&  5670 $\pm$     8&  39 $\pm$  16&Y&A2200-2909 & 2.32&20.60& 5658 \\ 
MARK817&  5779 $\pm$     7&  34 $\pm$  13&N&14293+5528 & 5.56&15.70& 5291 \\ 
$*$MARK501&  5990 $\pm$     7&  55 $\pm$  46&Y&16310+3727 & 8.50&16.00& 5309 \\ 
MARK817&  6241 $\pm$     6&  37 $\pm$   8&N&A1414+6042 & 5.12&15.50& 6595 \\ 
FAIRALL9&  6255 $\pm$     9&  22 $\pm$   9&N&A0105-5820 & 3.32&15.50& 5914 \\ 
$*$MARK335&  6266 $\pm$     6& 130 $\pm$  14&Y&00036+1928 & 0.82&16.34& 5936 \\ 
MARK279&  6360 $\pm$     7&  58 $\pm$   7&N&14010+6943 & 1.28&13.90& 6293 \\ 
MARK279&  6433 $\pm$     6&  40 $\pm$   7&N&14010+6943 & 1.28&13.90& 6293 \\ 
AKN120&  6750 $\pm$     7&  53 $\pm$  13&N&05071-0018 & 2.84&16.00& 6961 \\ 
MARK279&  6913 $\pm$     5&  65 $\pm$   8&N&A1401+7005 & 3.04&\dots& 6430 \\ 
MARK290&  7071 $\pm$     8&  23 $\pm$  10&N&A1505+6310 &10.73&\dots& 6584 \\ 
FAIRALL9&  7112 $\pm$     8&  32 $\pm$  10&N&A0106-5837 & 6.76&16.00& 7796 \\ 
MARK817&  7316 $\pm$    13&  17 $\pm$   7&N&I1049      & 6.91&12.70& 6728 \\ 
H1821+643&  7325 $\pm$     5& 298 $\pm$  20&Y&Tripp1     & 0.87&16.00& 7207 \\ 
MARK290&  7349 $\pm$     6&  21 $\pm$   7&N&A1505+6310 &12.51&17.00& 6584 \\ 
$*$MARK501&  7521 $\pm$    17&  53 $\pm$  36&Y&17048+4107E& 5.31&17.50& 7852 \\ 
H1821+643&  7537 $\pm$     9&  50 $\pm$  24&Y&Tripp1     & 0.97&15.00& 7207 \\ 
MARK279&  7767 $\pm$     6&  48 $\pm$   9&N&N5607      & 6.00&15.20& 7595 \\ 
MARK817&  7785 $\pm$    11&  28 $\pm$   9&N&14212+6010 & 7.57&15.60& 8520 \\ 
MARK279&  7845 $\pm$    10&  21 $\pm$  10&N&N5607      & 6.00&14.90& 7595 \\ 
H1821+643&  7853 $\pm$     7&  40 $\pm$  17&Y&18197+6340E& 1.34&\dots& 7756 \\ 
$*$3C273&  7870 $\pm$    11&  33 $\pm$  18&Y&A1228+0205 & 0.94&15.50& 7593 \\ 
AKN120&  7884 $\pm$    21&  20 $\pm$  25&N&A0509-0045 & 2.32&15.50& 7888 \\ 
H1821+643&  7941 $\pm$     8&  68 $\pm$  38&Y&Tripp2     & 0.34&15.60& 8250 \\ 
AKN120&  7977 $\pm$     5& 147 $\pm$  22&N&A0509-0045 & 2.32&15.60& 7888 \\ 
AKN120&  8037 $\pm$     7&  65 $\pm$  17&N&A0509-0045 & 2.32&15.60& 7888 \\ 
$*$MARK501&  8748 $\pm$     7&  66 $\pm$  57&N&16572+4012 & 2.22&15.60& 8714 \\ 
$*$3C273&  8830 $\pm$     8& 114 $\pm$  25&Y&A1226+0211 & 0.40&16.00& 8807 \\ 
Q1230+0115&  9239 $\pm$     8& 301 $\pm$  49&N&A1227+0144 & 0.52&14.80& 9281 \\ 
FAIRALL9&  9498 $\pm$     6&  84 $\pm$  13&N&A0114-6153 & 8.61&15.30& 8818 \\ 
$*$3C273&  9830 $\pm$    17&  46 $\pm$  22&Y&A1225+0223 & 0.39&16.70& 9752 \\ 
FAIRALL9& 11650 $\pm$     7&  16 $\pm$   8&N&A0101-5726 & 9.44&16.70&11114 \\ 
FAIRALL9& 11930 $\pm$    10&  22 $\pm$   9&N&A0120-5934 &10.10&13.83&12927 \\ 
FAIRALL9& 12099 $\pm$    18&  30 $\pm$  28&N&A0120-5934 & 7.73&14.39&12927 \\ 
FAIRALL9& 12202 $\pm$     6&  28 $\pm$   7&N&A0120-5934 & 6.30&14.39&12927 \\ 
H1821+643& 12317 $\pm$     6&  64 $\pm$  15&Y&18421+6358 & 6.94&14.39&12369 \\ 
FAIRALL9& 12416 $\pm$    29&  19 $\pm$  23&N&A0120-5934 & 3.47&14.39&12927 \\ 
PKS2155-304& 13589 $\pm$     6& 101 $\pm$  18&N&A2143-2933 &10.15&\dots&14030 \\ 
$*$3C273& 14689 $\pm$     7& 140 $\pm$  25&Y&A1224+0230B& 1.90&15.65&14860 \\ 
$*$3C273& 14984 $\pm$    14&  46 $\pm$  22&Y&A1224+0230B& 1.90&16.15&14860 \\ 
$*$3C273& 15239 $\pm$    33&  52 $\pm$  40&Y&A1224+0230B& 2.21&16.15&14860 \\ 
$*$3C273& 15928 $\pm$    19&  64 $\pm$  33&Y&A1223+0238 & 9.81&\dots&14986 \\ 
PKS2155-304& 16201 $\pm$     5& 346 $\pm$  23&N&A2155-3033A& 0.58&\dots&16200 \\ 
PKS2155-304& 16322 $\pm$    15&  62 $\pm$  34&Y&A2155-3033A& 0.58&\dots&16200 \\ 
PKS2155-304& 16922 $\pm$     7&  43 $\pm$  37&Y&A2155-3033B& 0.44&\dots&17093 \\ 
PKS2155-304& 16970 $\pm$     9& 389 $\pm$  68&Y&A2155-3033B& 0.44&16.58&17093 \\ 
PKS2155-304& 17116 $\pm$     9& 448 $\pm$  79&Y&A2155-3033B& 0.44&16.30&17093 \\ 
PKS2155-304& 17710 $\pm$     6& 139 $\pm$  21&N&A2156-3017 & 3.39&15.30&17179 \\ 
PKS2155-304& 18071 $\pm$     7&  99 $\pm$  20&N&A2156-2900 & 6.50&15.30&17739 \\ 
$*$3C273& 18270 $\pm$    24&  47 $\pm$  28&Y&A1229+0222 & 5.98&15.30&18889 \\ 
$*$3C273& 19031 $\pm$    12&  47 $\pm$  22&Y&A1229+0222 & 3.91&18.30&18889 \\ 
$*$3C273& 19953 $\pm$     6& 297 $\pm$  25&Y&A1222+0249 & 8.05&17.69&20661
\enddata
\tablecomments{A $*$ preceeding the sightline name indicates a pre-COSTAR HST observation.}
\tablenotetext{a}{Rest-frame equivalent width.}
\tablenotetext{b}{Y if the absorber is included in our \sample sample, N if not included.}
\tablenotetext{c}{Euclidian absorber-galaxy distance assuming $H_0$=\hseventy~70\kmsno\perMpcno.}
\tablenotetext{d}{Zwicky (blue) apparent magnitude of the nearest galaxy. For galaxies named Tripp\#, the POSS II J-magnitude is given,
which is roughly equivalent to $m_B$.}
\end{deluxetable}

\clearpage
\section{Galaxy - Absorber Relationships}\label{sec:assoc}
\subsection{Nearest-Neighbor Distributions}\label{sec:nn}
The lower panel of Figure~\ref{NNhist} presents the nearest-neighbor galaxy distances for 
the \Nsample~absorbers in the \sample sample.
In this histogram, the stronger \lya features, are displayed in grey bins,  while the weaker
features,  W\lt68\mang, are displayed in dark bins.  The choice of the 68\Mang 
cutoff divides our restricted \sample sample equally by number and corresponds to $\logNhcm=13.15-13.18$
assuming a Doppler parameter of $b=20-30\kms$ (see Papers~I and~II). 
A two-sided K-S test indicates that the stronger and weaker features are drawn from different populations 
at the 99.3\% ($\gt3\signo$) confidence level, similar to the result obtained for the
median split \definite\ sample shown in the upper panel of Figure~\ref{NNhist}.
This also confirms our earlier contention \citep{Stocke95} that the weak and 
strong \lya features are distributed differently with respect to
nearest neighbors, with the stronger features generally having closer 
nearest galaxy neighbors. Eighty percent of our stronger \sample absorbers have a nearest
galaxy neighbor within 1\hsfi Mpc, while only 45\% of the weaker \sample absorbers have a known
galaxy within 1\hsfi Mpc.

Note that there is considerable spread in the nearest-neighbor distances in both the
strong and weak absorber samples. If we define being in a ``void'' as 
lacking a nearest neighbor within 3 \hsfi Mpc, then 8 (17\%) of our 
\sample absorbers are in voids and 38 (83\%) are not. 
Our definition of ``void" is not entirely arbitrary, in that 3
\hsfi Mpc  is the median distance from a random point in the \lowz Universe 
to the nearest galaxy (see \S~\ref{sec:CDF} below). 
In our full sample, 32 of 81 of our \foursig \lya absorbers (40\%) lie in voids according
to the above definition, illustrating the need to restrict the sample region according to
limiting magnitude. While our use of a  ``retarded Hubble flow'' model could create an 
underestimation of the void fraction, we have used other ``void'' definitions guided by the
\citet{Slezak93} wavelet technique for identifying voids. This secondary methodology gives
nearly identical results to the 3\hsfi Mpc retarded Hubble flow definition.
Interestingly, even if we exclude the 17 local supercluster absorbers, there are only 3 
out of 17 absorbers in voids at $cz<10,000\kms$ but 5 out of 12 at greater \cz.
The reason for this possible difference is not clear but maybe due either to the different 
galaxy survey methodology employed in these two redshift regimes (i.e., ``pencil-beam''
surveys at higher \cz, CfA redshift survey at lower \cz) or to the greater sensitivity of
our GHRS spectra at higher \cz, causing statistically weaker absorbers to be 
detected at higher \cz. 
 
Of these 8 ``void'' absorbers, only two are in the strong half of the sample ($\Wno\gt68\mang$):
\begin{enumerate}{
\item MRK~509  : nearest galaxy = 4.7 \hsfi Mpc  ; $\Wno = 209$\Mang  ; $\cz =2549\kmsno$ \\
	(galaxy survey complete to 0.2\Lstar at this location)
\item 3C~273  : nearest galaxy = 8.0 \hsfi Mpc  ; $\Wno = 297$\Mang ; $\cz =19,953\kmsno$ \\
(galaxy survey complete to 0.3\Lstar at this location)
}\end{enumerate}
Because of its proximity, the \cz=2549\kms absorber of MRK~509 
is our best candidate for studying \lya absorbers in \lowz galaxy voids. 

Even though the \sample sample in Table~\ref{linelist_sorted_def} is defined by having
galaxy redshift survey data complete to at least \Lstarno, there is still a possible 
distance-dependent bias in
nearest-neighbor distances because lower \z\ locations within the same survey region 
are more sensitive to lower luminosity galaxies. Absorbers at $cz_{abs}<cz_c$ 
limit in Table~\ref{completeness} have been
surveyed to below \Lstar luminosities and so these absorber regions have fainter known galaxies. 
Thus, a distance dependent bias could be introduced into the \sample sample, as a result.
Using their Virgo region sample, IPF99 showed that decreasing the
luminosity limit of the galaxies surveyed does decrease the nearest-neighbor distance to some of their 11
absorbers, so that such an effect could be present in the \sample sample as well. The use of different types of
surveys (i.e., wide-angle, shallow surveys, deep pencil-beam surveys and limited velocity range 21~cm
surveys) might also create some bias. For example, the closest galaxy-absorber pair in
the \sample sample is due
to a galaxy at $\Dperp=96\hsfi $kpc from the $cz=1950\kms$ absorber in the MRK~335
sightline, which was discovered in a pointed \hone survey \citep{VG96}. This
galaxy at $M_R\sim-15$ (0.04\Lstarno) and $M_{HI}=4\times10^7M_{\Sun}$ is below the CfA
redshift
survey limit at that distance ($M_B=-16.7$) and so is the least luminous galaxy in any
galaxy redshift catalog used herein.

\begin{figure}[htb]
\epsscale{0.8}
  \plotone{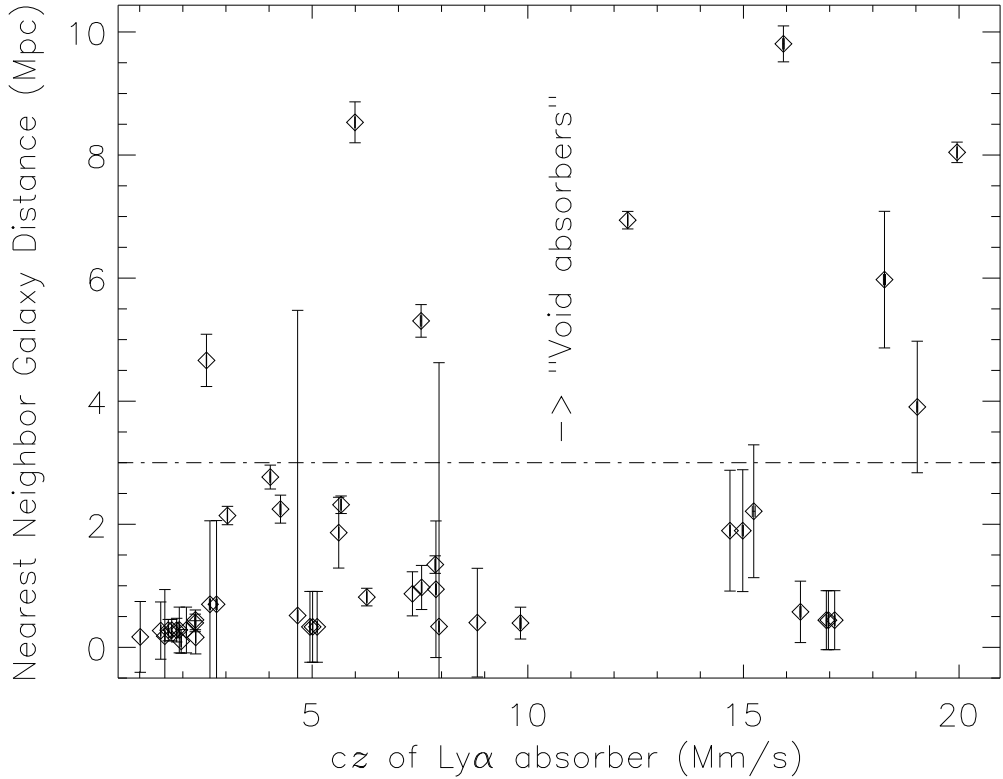}  
  \caption{\label{NNdistczsample} 
Nearest galaxy distances (\Dtotno) for all absorbers in the \sample sample, in \hsfi Mpc.
An absorber-galaxy ``coincidence'' is defined as any galaxy within 350 \hsfi kpc of 
an absorber using our ``retarded Hubble flow'' model (see \S~\ref{sec:non}). 
Eight ``void absorbers'' have $\Dtotno \ge 3$ Mpc.}
\end{figure}

In Figure~\ref{NNdistczsample} we show the nearest galaxy distances to absorbers in the
\sample sample as a function of recession velocity. We discount the absorbers at
$cz\leq2000\kmsno$, almost all of which have uncommonly close absorber-galaxy distances.
These absorbers are nearly exclusively in the 3C~273 and Q~1230+0115 sightlines which
penetrate the southern region of the Virgo cluster at $\cz < 2000\kmsno$.
This figure shows that there is
no evident correlation between absorber distance and recession velocity (correlation
coefficient=0.45). Similarly, the mean distance to the nearest 3 galaxies is not strongly
correlated with redshift (correlation coefficient=0.59). Are the small
absorber-galaxy distances at $cz\leq2000\kmsno$ a result of the lower luminosity limits for galaxies, 
or of a larger space density of galaxies in the Virgo region? The answer is not
obvious based upon the present information, but to be safe we investigate the
\sample 
sample, with and without the Virgo absorber-galaxy pairs. Despite the absence
of a correlation in Figure~\ref{NNdistczsample}, we devised comparison methods 
that substantially reduce, if not eliminate, 
the effect of any distance-dependent biases by using the same CfA redshift survey
volume to measure the statistics of the comparison samples (see \S~\ref{sec:CDF}). 
\subsection{Cumulative Distribution Functions}\label{sec:CDF}
In order to achieve an unbiased comparison to our \sample sample of absorbers,
we determine: (1) the galaxy-galaxy nearest-neighbor distribution and (2) the
random location-nearest-neighbor galaxy distribution of the \about5200
CfA galaxies in the first six CfA slices, that dominates our survey volume. 
Specifically, the distributions of
nearest-galaxy-neighbor distances were 
calculated using the region covered by the first six CfA slices:
$$ 8.25^h \lt \alpha \le 16.75^h ~~; ~~3.\degr5 \lt \delta 
   \lt 43.\degr5~~;~~ 575 \lt cz \lt 7230\kmsno~.$$
The velocity constraints correspond to our lower spectral limit (1218\ang) 
and the \Lstar completeness limit. To avoid any
possible ``edge'' effects, we consider only galaxies greater than \about0.\degr5 from the edge of 
the ``well-surveyed'' regions defined in \S~\ref{sec:sample}.
We have compared our absorbers to a random distribution based upon 250 sightlines 
through the above region that are consistent with the varying sensitivity, 
\sensno, and wavelength coverage of
our spectra. The random absorber location sample is simulated identically 
as in Paper~II; for each pixel of each random spectrum (sightline), 
we compute the probability of a \lya absorber 
occurring given our $d\mathcal{N}/d\W$ selection function.  We combine this probability
with the signal to noise (S/N) of each spectral location, 
to determine the probability, $P(\lambda)$, that a randomly placed 
absorber will occur at that location. If $P(\lambda)$ is greater than a 
uniformly distributed random number, then we locate an absorber  
at that pixel. As with the TPCF, because 
of our limited spectral resolution, two
absorbers are not allowed to occur within 50\kms of one another. 

\begin{figure}[htb]
  \plotone{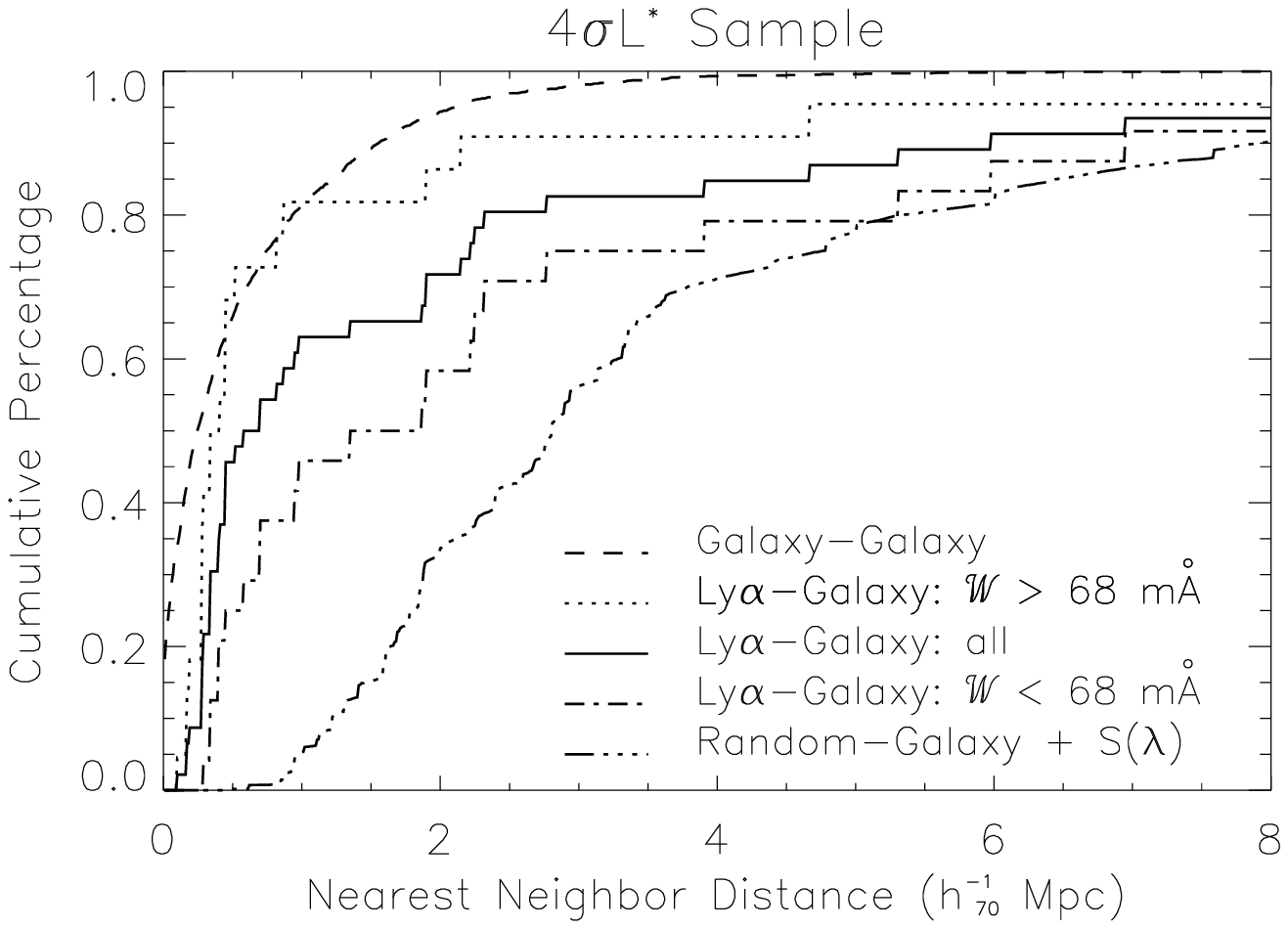}
  \caption{\label{NN_gcum} Nearest-neighbor cumulative distribution function (CDF)
for our \sample sample (solid line, \lya-galaxies). The dashed line indicates 
the CDF for the galaxy-galaxy nearest-neighbor distances in well-surveyed regions of the CfA redshift 
survey (see text). This curve shows the strong clustering
of galaxies; 90\% of the galaxies have nearest-neighbors within 
2 \hsfi Mpc. The triple dot-dashed line indicates the distribution
expected for random locations within our survey volume accounting for our wavelength and sensitivity limits, \sensno, combined with 
our \dndW\ distribution. Dotted and dot-dashed lines show the CDF
 for our \sample sample when split above and below \W = 68\mang, 
the median of the \Wno-distribution. 
Fifty percent of randomly distributed locations are $\ge$ 3\hsfi Mpc from the nearest CfA galaxy, 
suggesting a definition of a void as being locations $\ge$ 3\hsfi Mpc from the nearest galaxy.}
\end{figure}

The galaxy-galaxy, random absorber-galaxy, and observed \lya absorber-galaxy cumulative 
 distribution functions (CDFs) are displayed for our \sample sample in  Figure~\ref{NN_gcum}.  
The clustering of galaxies is revealed by the dashed line, which
shows that  $\gt 90$\% of the galaxies in the \sample region are within 
2 \hsfi Mpc of another galaxy. For a random distribution (triple dot-dashed line) 
that matches the selection function of our \sample sightlines, only 27\% are within 2 \hsfi Mpc 
and 50\% are within 3 \hsfi Mpc of the nearest galaxy. The solid line shows our \sample absorber sample.
The inclusion of the few absorbers at $cz\leq2000\kmsno$, that may be biased by being in the Virgo region 
(see Figure~\ref{NNdistczsample}), does not alter the  result significantly. It is clear that
our \sample \lya sample does not cluster with galaxies as strongly as galaxies cluster, 
but they are more clustered than a random distribution. K-S tests give only a \Exp{1}{-5} probability that our \sample 
absorber sample clusters with galaxies as strongly as galaxies cluster with other 
galaxies, and a \Exp{9}{-4} probability that the \sample absorber sample is randomly distributed 
with respect to galaxies.  
Figure~\ref{NN_gcum} also shows that, while the median distance between \Lstar galaxies
is \about250\hsfi kpc, the median \lya cloud-galaxy distance is \about500\hsfi kpc. Because
\lya absorbers are typically twice as far away from bright galaxies as these bright
galaxies are separated, it is problematic to associate these clouds with individual
bright galaxy halos.
The results shown in Figure~\ref{NN_gcum} are similar to those in
\citet{Morris93}, \citet{Stocke95}, \citet{Tripp98a}, and IPF99, although the
different techniques used do not allow a more quantitative comparison with these earlier
results. It has been suggested \citep{Impey97b} that,
because \lya absorbers correlate with galaxies less strongly than
galaxies correlate with galaxies, their clustering is similar to
that of low-surface-brightness (LSB) galaxies. \citet{Impey97b} take
one further logical step and associate \lya clouds with LSB  galaxies. 
This speculation remains an open question. 
We note that no LSB galaxies have been found  close to any low column density \lya clouds to our knowledge, including
the pair of Virgo absorbers towards 3C~273, where sensitive limits have been set on their absence \citep{Rauch96}.

To further explore the possibility that the weaker and stronger 
 \lya absorbers sample different populations, we split the
\sample sample into two distributions at \W above and below 68\mang. 
The CDFs for these split samples are also shown in Figure~\ref{NN_gcum}.
It is clear that the stronger absorbers cluster more like galaxies, and the
weaker absorbers appear to cluster more randomly, although neither 
sample is well-matched by the comparison samples. The K-S probability that
the strong absorber-galaxy and the galaxy-galaxy distributions are 
drawn from the same parent distribution is 0.3\% (\foursig difference). The K-S probability that
the weak absorber-galaxy and random-galaxy distributions are 
drawn from the same parent distribution is only \Exp{4}{-4}. Additionally, 
a K-S test indicates that the weak and strong \lya samples differ at 
the 99.1\% (\threesigno) confidence level. These results confirm our earlier suggestion
\citep{Stocke95} of such a dichotomy, which was based on fewer absorbers.

The primary difference between the galaxy-galaxy and
the strong absorber-galaxy nearest-neighbor
distributions in Figure~\ref{NN_gcum} is due to the two absorbers 
listed in \S3.1. These two ``void absorbers'' are several \hsfi Mpc away from the nearest known galaxy. In order to
have these two distributions match each other statistically, very faint galaxies would
have to be within $1-2\hsfi $Mpc of these absorbers. 
If these two ``void absorber'' are removed from the strong absorber sample, this sample still
deviates from the galaxy-galaxy CDF at the 98.4\% ($>$\threesigno) level.
The very nearby ``void absorbers'' in the MRK~509 and MRK~421 sightlines are in regions already 
surveyed to 0.2\Lstar and 0.3\Lstar by the CfA survey data, respectively. 
Thus, the evidence is already strong that at least some higher column density absorbers 
are not physically associated with galaxies. 
New deeper galaxy surveys in the optical \citep{McLin02} and \hone \citep{Hibbard02} 
extend this result to impressively low limits for these and other void absorbers.
The absence of metal absorptions in the strong
\lya systems found in the \PKS\ sightline at $cz\sim17,000\kms$ \citep{pks} is further evidence that 
some higher column density \lya absorbers are not ``recycled gas'' in extended galaxy halos. 
\begin{figure}[htb]
  \plotone{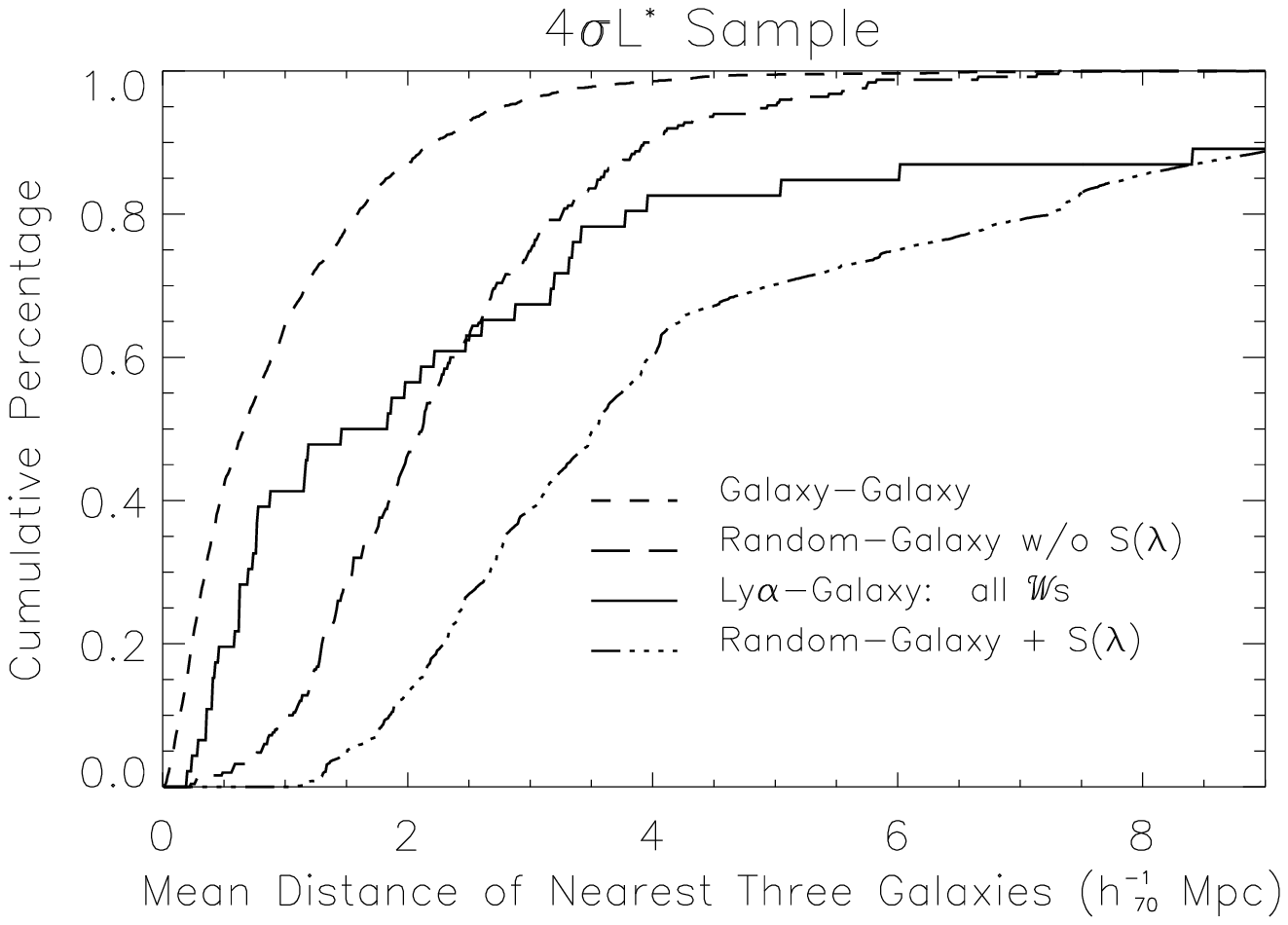}
  \caption{\label{NN_gcum_m3}
Nearest-neighbor cumulative  distribution function (CDF) using the mean distance to the nearest three known galaxies.
The short dashed line indicates the CDF for the galaxy-galaxy distances,
the solid line gives the CDF for our \sample sample,
and the dot-dashed and dashed lines show the CDF for random distribution
with, and without, the inclusion of our sensitivity function, \sensno.}
\end{figure} 

\citet[GG98]{Grogin98} recently used a subset of our \sample sample,
 in conjunction with the CfA redshift survey, to investigate the relationship between
galaxies and \lya clouds in a somewhat more robust manner.
As shown by IPF99, the exact nearest-neighbor distance for each absorber depends upon the absolute
magnitude limit. 
GG98 determined a galaxy density at each point within the CfA survey region by
averaging over the 1--3\hsfi Mpc regions,
using the detailed selection function of the survey to determine an 
averaged galaxy density at the location of each \lya cloud. By doing this, 
GG98 avoided using just the nearest-neighbor galaxy distance as an indication of \lya absorber
environment. They also used the CfA survey mean luminosity function to correct for
the presence of galaxies below the survey limit at each absorber. From
 this procedure, they found that \lya clouds are located randomly with respect to 
the averaged galaxy density. We make the following comments about this work: 
\begin{enumerate}

\item Although GG98 claim that their result is insensitive to their 
   smoothing lengths, these lengths are comparable to or larger than
   nearest-neighbor distances for 80\% of our sample. 

\item The GG98 sample contains roughly equal numbers of stronger and 
   weaker \lya absorbers by the definition of Figure~\ref{NN_gcum}.

\item Because the sample used by GG98 contains 5 absorbers that are no 
   longer in our \sample sample (e.g., the \cz =2060\kms absorber in 
   the 3C~273 sightline), this study should be revised with newer 
   absorber data.

\item The GG98 analysis technique does not take into account the varying 
   \W sensitivity of the HST spectra, as we have 
   done here.

\end{enumerate}

In order to use the \sample sample to determine whether our current results are 
consistent with the GG98 analysis, we have repeated
the nearest-neighbor analysis using the mean distance to the nearest 3 galaxies (see Table 10 of Paper~I). 
We have generated galaxy-galaxy
and random location-galaxy data to compare with the \lyano-galaxy data in a manner identical to the
descriptions above. The use of the mean of the nearest three galaxy distances provides some smoothing of the
galaxy environments, but on a scale more local to the \lya absorbers than the GG98 analysis.
Our results  are shown in Figure~\ref{NN_gcum_m3}, a CDF similar to Figure~\ref{NN_gcum}. To illustrate the importance
of using the spectral sensitivity function, \sensno, when calculating the random CDF,
we have also included the random CDF without using \sensno. As mentioned above,
without considering \sensno, our \lya CDF appears more random than it truly is. 
This suggests that, by not including the sensitivity of the \lya survey,
GG98 may have drawn the incorrect conclusion that \lya absorbers are randomly
distributed with respect to the galaxy density.
New work is underway to utilize the GG98 technique 
on a much larger sample of absorbers, incorporating our \sens to revisit these results \citep{Grogin02}.
\subsection{\lya Absorber Searches near Galaxies} \label{sec:non}
Because of the suggestion by L95 that extended galaxy halos are the sources of \lya 
absorption, we have searched for absorbers near known galaxies along 
our sightlines. For this purpose, we classify galaxy-sightline and galaxy-absorber 
matches into two categories: ``coincidences'' and ``non-detections''. 

We define a ``coincidence'' as a \real\ absorber within 
350 \hsfi kpc of a galaxy using our ``retarded Hubble flow'' model.
This distance was chosen to be inclusive of all ``coincidences'',
based upon an earlier analysis of higher \W absorbers by L95,
but only five of these are in as close proximity as the ``associations'' found by L95
(our impact parameter limit is 1.5 times the limits of galaxy halos suggested by L95; 230\hsfi kpc).
An absorber-galaxy coincidence does not necessarily imply that the  
absorber is gravitationally bound or physically connected with a galaxy
in any way;  for this present analysis it only implies proximity. 
In our full \foursig sample, we find that 14 of our \Nabs\ absorbers 
(17\%) are ``coincident'' with 11 different galaxies.
All of these ``coincidences'' involve absorbers in the \sample sample.
Several galaxies are ``coincident'' with absorbers in both the 3C~273 and Q~1230+0115 sightlines
up to \cz\about 2300\kmsno, although in most of these cases the velocity difference between
the galaxy and the \lya absorber  in at least one of these sightlines is quite large (Table~\ref{detections}).
Fully 2/3 of the ``coincidences'' are in the Virgo region of the 3C~273/Q~1230+0115 sightlines,
suggesting that these matches are, in fact, truly coincidental; 
most occur in the region of highest galaxy density in our surveyed volume.

Table~\ref{detections} lists,  in order of increasing separation distance, 
galaxy-absorber information for the ``coincidences''.
Eleven galaxies are listed for 21 ``coincidences''.
For brevity, Table~\ref{detections} lists a maximum of three associated galaxies for each absorber (complete
information regarding these nearest galaxies can be found  in Table~10	of Paper~I). 
If only the nearest galaxy is listed, the number of ``coincidences'' drops to 16.
Galaxy information is derived from
the CfA survey, except where previously noted. Information in Table~\ref{detections} includes by
column: 
(1) sightline target name,
(2) galaxy name,
(3) the galaxy Zwicky magnitude, 
(4) the galaxy heliocentric radial velocity in\kmsno,
(5) the galaxy-absorber 3D separation distance (\Dtotno) in \hsfi kpc assuming our retarded Hubble flow model,
(6) the heliocentric radial velocity in\kms for the associated \lya absorber,
(7) the rest-frame equivalent width (\Wno) of the \lya absorber in \mang, 
(8) the significance level (SL) of the \lya absorber, and 
(9) whether the absorber is in our \sample sample. 
Because the \Dtot values assume a retarded Hubble flow model, these 3D
distances should be considered lower limits. 

As previously mentioned, and indicated in Table~\ref{detections}, our closest galaxy-absorber 
pair is related to the 1963\kms absorber in the MRK~335 sightline. 
The galaxy impact parameter at the velocity of the galaxy (1950\kmsno) 
is \Dperp = 96\hsfi kpc. The ``coincident'' galaxy, detected by a 21~cm 
survey  \citep{VG96}, has an estimated $M_R= - 15$ (0.04 \Lstarno) 
and an \hone mass \about \Exp{4}{7}\Msolar. 

Almost all of the  ``coincidences'' are in the well-surveyed Virgo cluster region surrounding the 3C~273
and Q~1230+0115 sightlines. These include the Haynes \& Giovanelli \hone cloud (HG cloud hereafter;
$cz_{gal}=1105$\kmsno),  which is $116\hsfi $kpc off the Q~1230+0115 
sightline but has no \lya detection in the new Q~1230+0115 E140M spectrum to $\le$ 20\Mang (\threesigno). 
However, there is a \lya detection in the 3C~273 spectrum at $cz_{abs}=1013\kmsno$, 
with an impact parameter (Table~\ref{3CQfield}), and 
\Dtot (Table~\ref{detections}) of $170\hsfi$ kpc from the HG cloud. 
Therefore, if the 3C~273 detection and the Q~1230+01115 non-detection are both probing the 
HG galaxy halo, that halo must be quite patchy or asymmetric.
\citet{Bowen96} came to a similar conclusion involving the covering factors of galaxy
``halos'' using independent data.
This is the only ``non-detection'' of a bright galaxy in our sample,  by which we mean that no $\gt
\threesig$ absorber is found within 350 \hsfi kpc of  a known galaxy assuming our retarded Hubble flow
model.
\begin{deluxetable}{lcrccccrc}
 \tablecaption{\label{detections} Galaxy-\lya ``Coincidences".}
 \tabletypesize{\scriptsize}
 \tablecolumns{9}
 \tablewidth{0pt}
 \tablehead{
\colhead{Target}&
\colhead{Galaxy} &
\colhead{$m_B$\tablenotemark{a}} & 
\colhead{$cz_{gal}$}  &
\colhead{D$_{tot}$\tablenotemark{b}} &  
\colhead{$cz_{abs}$} &
\colhead{\Wno} &
\colhead{SL} &
\colhead{In \sample}\\
\colhead{}&
\colhead{Name}&
\colhead{}&
\colhead{(\nokmsno)}    & 
\colhead{(kpc)}   &  
\colhead{(\nokmsno)}  &
\colhead{(\nomang)} &
\colhead{}&
\colhead{sample ?}
} 
\startdata
MARK335&00025+1956 & 16.0&        1950&          96&        1961&         229& 28.1&Y\\
Q1230+0115&A1228+0116 & 18.3&        2289&         162&        2298&         438& 23.9&Y\\
3C273&12285+0157 & 15.6&        1105&         169&        1012&         369& 34.8&Y\\
3C273&A1225+0152 & 16.3&        1298&         181&        1012&         369& 34.8&Y\\
3C273&A1225+0152 & 16.3&        1298&         181&        1583&         372& 42.0&Y\\
3C273&A1224+0146 &  NA&        1275&         231&        1012&         369& 34.8&Y\\
3C273&12277+0254 & 15.3&        1635&         260&        1583&         372& 42.0&Y\\
3C273&A1224+0146 & NA&        1275&         261&        1583&         372& 42.0&Y\\
Q1230+0115&A1225+0152 & 16.3&        1298&         271&        1487&         138&  6.3&Y\\
Q1230+0115&12308+0147 & 15.3&        1643&         279&        1487&         138&  6.3&Y\\
Q1230+0115&12308+0147 & 15.3&        1643&         279&        1742&         241& 10.9&Y\\
Q1230+0115&12308+0147 & 15.3&        1643&         279&        1663&         384& 16.9&Y\\
Q1230+0115&12308+0147 & 15.3&        1643&         279&        1857&         141&  6.6&Y\\
MARK817&14332+5934 & 15.5&        1920&         281&        2082&         135& 25.3&Y\\
MARK817&14332+5934 & 15.5&        1920&         281&        1918&          28&  5.3&Y\\
Q1230+0115&12270+0107 & 17.0&        2234&         292&        2298&         438& 23.9&Y\\
Q1230+0115&A1224+0146 &  NA&        1275&         293&        1487&         138&  6.3&Y\\
PKS2155-304&A2157-3025 & 14.4&        5187&         332&        4948&          64& 14.3&Y\\
PKS2155-304&A2157-3025 & 14.4&        5187&         332&        5010&          81& 18.1&Y\\
PKS2155-304&A2157-3025 & 14.4&        5187&         332&        5115&         218& 48.1&Y\\
H1821+643&Tripp2\tablenotemark{c}    & 20.6&        8250&         337&        7940&          67&  9.4&Y\enddata
\tablecomments{An ``concidence'' is defined as any galaxy within 350\hsfi
kpc of one of our sightlines that coincides with a \real\ \lya absorber
assuming our ``retarded Hubble flow" model.  
Some galaxies are associated with multiple
\lya absorbers, and some absorbers have multiple associated galaxies. We limit the number of
``coincidences'' per absorber to three.}
\tablenotetext{a}{Blue (Zwicky) magnitude, where available.}
\tablenotetext{b}{Total Euclidean distance (in \hsfi Mpc) to the nearest absorber
using our ``retarded Hubble flow" model.}
\tablenotetext{c}{The Tripp2 galaxy is taken from \citet{Tripp98a}.}
\end{deluxetable} 

\begin{figure}[htb]
  \plotone{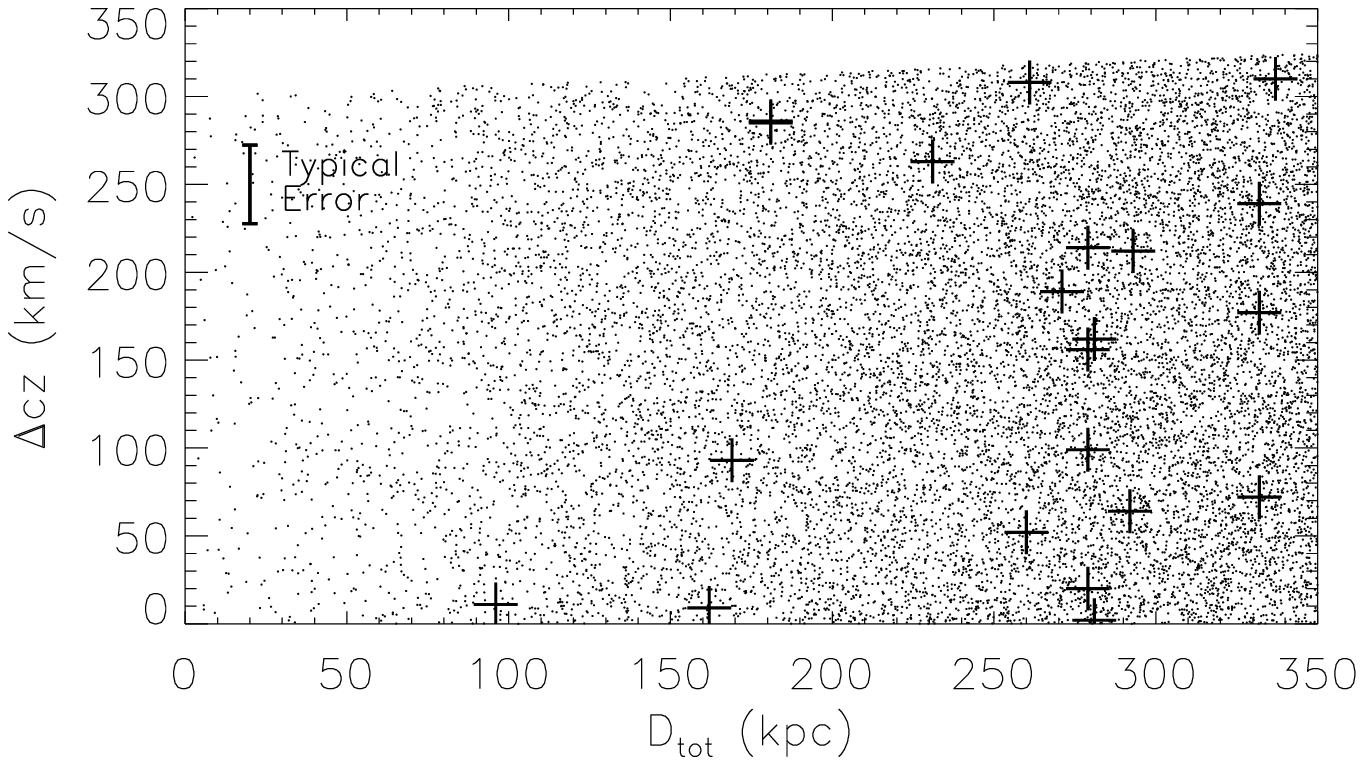}
  \caption{\label{figure5}
Galaxy-\lya velocity difference (\Dv) versus total, retarded Hubble flow, 3D distance (\Dtotno)
between galaxy and absorber as presented in Table~\ref{detections} (the '+'s).  
A random distribution of galaxies locations relative to the absorbers is shown by the numerous dots. 
Our observed distribution is consistent with a random distribution at the 88\% confidence level.}
\end{figure}

In Figure~\ref{figure5}, the solid `+'s present the \Dv\  vs \Dtot results given in Table~\ref{detections} for
all detected \lya absorbers within 350\hsfi kpc of a known galaxy.  
Also displayed, by the numerous dots, is 
the distribution expected for random galaxy-\lya absorber coincidences using our retarded Hubble flow model. 
Due to our retarded Hubble flow 3D distance determination (\Dtotno), if the difference in velocity between
the absorber and the nearest galaxy were $\gt 300$\kmsno, the computed \Dtot would become very large. 
The upper boundary of `+'s and dots in Figure~\ref{figure5} in due to this assumption.
As implied by Figure~\ref{figure5}, and confirmed by K-S tests, the observed 
distribution of ``coincidences" shows no significant departure from random galaxy-absorber locations. 
K-S tests give an 88\% probability that the \Dv\ distribution of \lya clouds with respect to the 
nearest galaxy is consistent with the random distribution. 
The \Dtot results are also consistent with random coincidences.
Therefore, our absorber-galaxy ``coincidences" are statistically consistent with chance 
galaxy-absorber locations and are
not indicative of any physical or causal relationship.
\subsection{Equivalent Width versus Impact Parameter Relationship}\label{sec:relationship} 
\begin{figure}[htb]
\epsscale{0.8}
  \plotone{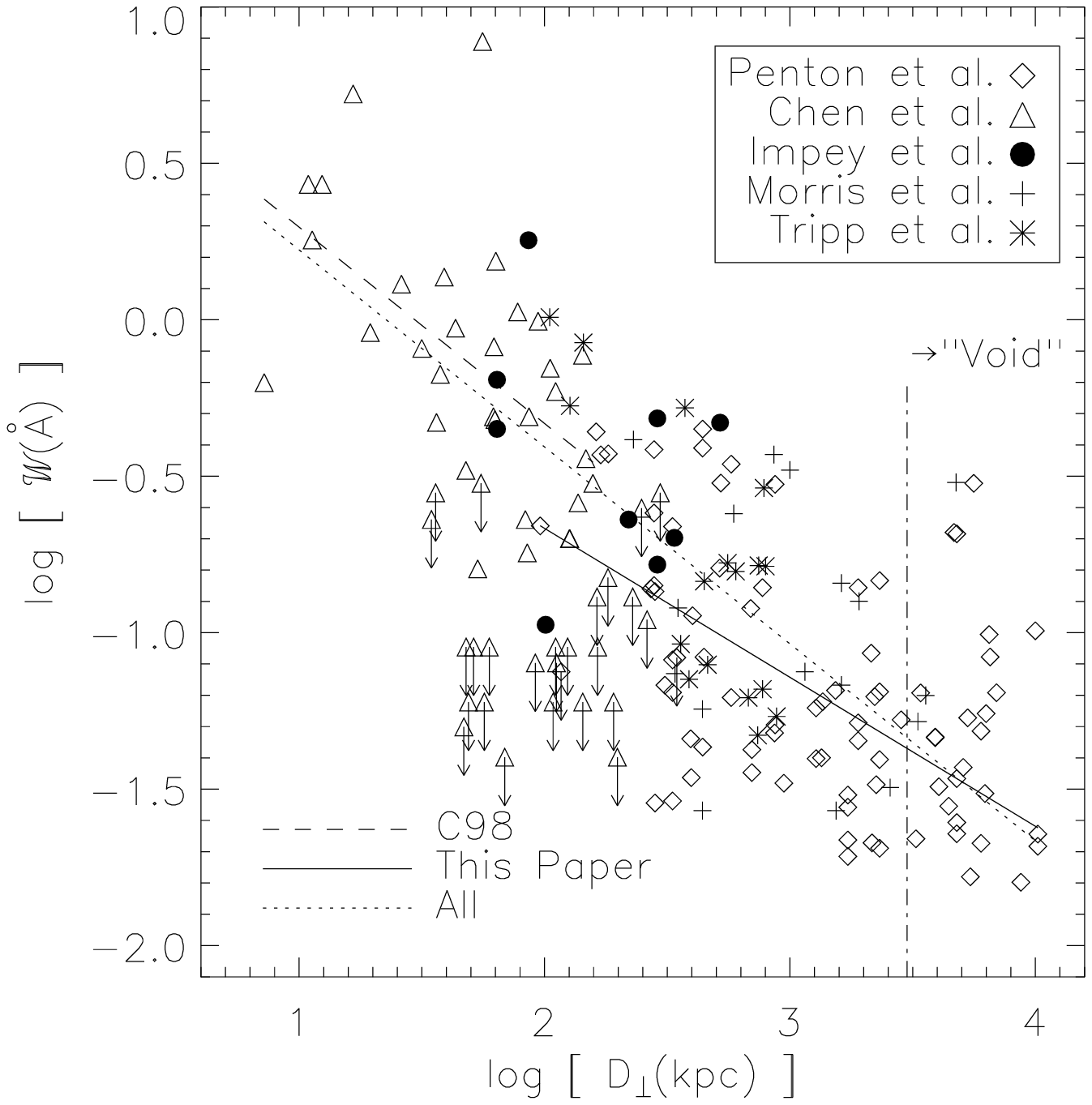}
  \caption{\label{Lanzetta} \lya absorber equivalent width (\Wno) 
versus impact parameter (\Dperp)  from the nearest galaxy to the line of sight.
Since the absorber could be at a different velocity than the nearest 
galaxy, \Dperp\ should be considered a lower limit to the actual 
galaxy-absorber distance. Non-detections are indicated by downward 
arrows, originating at the \threesig\W upper limit.  All distances 
are calculated assuming $H_0$  = 70\kms\perMpcno. The dashed vertical 
line indicates our ``void'' definition, \Dperp \gt\ 3\hsfi Mpc.
The dashed linear fit corresponds to the C98 data only (triangles) 
and returns a high probability of correctly modeling those data 
(\gt 99.99\%). The dotted line, with a slope of -0.57, corresponds to a linear fit of all data 
and returns a low probability of correctly modeling all data (\lt 0.01\%). 
The solid line represents a linear fit to our \sample data alone 
and also returns a low probability of correctly modeling our data 
(\lt 8\%). } 
\end{figure}

\citet{Lanzetta95}, and more recently C98, used HST/FOS spectra of
bright quasars observed by the Key Project Team \citep{Bahcall93} 
to study the association of nearest galaxies to \lya clouds at 
0.1\lt\z\lt0.3. C98 reported a distinct
anticorrelation between absorber \W and impact parameter to the LOS 
(\Dperp) out to 230 \hsfi kpc, which led them to suggest
that most galaxies are surrounded by extended gaseous halos of 
\about 230 \hsfi kpc and that most or all \lya absorptions arise 
from these extended galaxy halos. HST/FOS data have wavelength resolution of \about 1\Ang
and minimum  detectable \Ws\ about 10 times greater than our HST/GHRS spectra. 
The FOS survey did contain some weaker absorbers, so there is some small overlap
in \W between the FOS and GHRS surveys.
The minimum \W used by L95/C98 is \about400\mang, roughly 6 times higher than the 
median \W studied herein. 
Because many or most of the L95/C98 \lya absorptions correspond to
metal-line systems and have substantially higher column density, the results
in Figure~\ref{Lanzetta} show that the L95/C98 conclusions may not be appropriate 
to extrapolate  to weaker \lya detections. 

At high-\z, it is still debated (Sargent 1987; Tytler 1987; Songaila \& Cowie 1996, Tytler
\eti\ 1995, Weymann 1995)  whether there are two distinct populations of \hone clouds: 
metal-poor absorbers that do not cluster and higher column density 
(\Nh \gt 10$^{14}$\percmtwono) metal-bearing absorbers that cluster 
like galaxies. \citet{Tripp98a} used GHRS/G140L+G160M spectra of H~1821+643 
(G160M included in our sample) and a GHRS/G140L spectrum of PG~1116+215
with minimum \W \about 50\Mang to examine the \W versus \Dperp\ 
relationship. They concluded that the anticorrelation
persists to 640 \hsfi kpc, but at a significantly larger dispersion
than for just the C98 data. In Figure~\ref{Lanzetta}, we collect 
\W and \Dperp\ from our study, C98, \citet{Morris93}, \citet{Tripp98a}, 
and IPF99. 
We also include our non-detections of \S~\ref{sec:non} and those of C98.

As indicated in Figure~\ref{NN_gcum}, it is rare to have a 
random sightline with an impact parameter from an $\sim\Lstar$ galaxy 
less than 1 \hsfi Mpc. Therefore, since FOS and G140L spectra 
cannot detect absorbers with $\Wno \lt\ 50$\mang, or 
$\log[\Wno(\noang)] \lt -1.3$, it not surprising that the lower left 
corner of Figure~\ref{Lanzetta} is sparsely populated using galaxy surveys 
of fields that do not go exceptionally faint (as in the L95 and C98 
studies). Galaxy detection surveys that extend significantly below \Lstar  
(van Gorkom et al. 1993, 1996; Shull \etl 1996; Rauch, Weymann \& Morris 1997; IPF99) would be expected 
to find smaller impact parameters, by chance in some cases (see e.g., \S~\ref{sec:non} and IPF99 Figure~13). 
The upper  right corner of Figure~\ref{Lanzetta}, with large impact parameters and high 
\Wnos, is also sparsely populated. Perhaps this is merely a consequence of the scarcity of
large-\W absorbers in  galaxy voids. Our GHRS survey does not have sufficient pathlength to 
detect such rare, strong absorbers, but the FOS Key Project does, should such
absorber systems exist. 
We do not know if any are present in the L95 and C98 samples, since, as pointed out by
\citet{Tripp98a}, neither L95 nor C98  plot the absorbers for which they failed to discover
nearby galaxies. We have not included the ``failed searches'' from C98
in Figure~\ref{Lanzetta}.  However, the lower limits implied by their non-detections place 2/3
of their absorbers to the right of the dash-dot vertical (``void'') line. 
We caution that the L95 \& C98 galaxy searches do not go very deep and could have left nearby
\Lstar galaxies undetected, to say nothing of dwarf or LSB galaxies.

In Figure~\ref{Lanzetta} we overplot a linear fit that corresponds to the C98 data only
and returns a high probability of correctly modeling their data alone 
(\gt 99.99\%). Even so, the C98 correlation is quite broad at any \W or 
\Dperp, so that the high correlation coefficient may merely reflect 
the biases mentioned above against finding galaxy-absorber pairs in the upper left
and lower right quadrants of Figure~\ref{Lanzetta}.
When all the data in Figure~\ref{Lanzetta} are fitted, the correlation coefficients 
indicate a low probability of correctly modeling the data (\lt 0.01\%). The solid 
line in Figure~\ref{Lanzetta} is a linear fit to our GHRS data alone and 
also returns a low probability of correctly modeling our data (\lt 8\%).

As seen in Figure~\ref{Lanzetta}, the region with \Dperp \gt~200 \hsfi kpc ($\log\left[ \Dperp \right] \gt 2.3$) and 
$\Wno \lt 400$\Mang ($\log\left[ \W(\noang) \right] \lt -0.4$) 
is a nearly uniform scatter plot.  Based upon the low probability for the fits 
to the low-\W absorbers in Figure~\ref{Lanzetta}, we conclude that the results of C98 and L95, 
the anticorrelation between galaxy impact parameter (\Dperp) and 
absorber \Wno, does not extend above \Dperp \gt 200 \hsfi kpc. In fact, 
when the non-detections of C98 are included (the triangles with
downward arrows in Figure~\ref{Lanzetta}), it appears that the 
reported anticorrelation may not extend above 
\Dperp \gt~50 \hsfi kpc ($\log\left[ \Dperp ({\rm kpc})\right] \gt 1.7$).
This distance may be the full extent of gaseous halo material physically  
associated with the nearby galaxy. Alternately, gaseous halos may extend somewhat further, but
with lower covering factor \citep[see][]{Mo00}. In addition, it has been demonstrated
\citep{Dave99,Linder00} that this anticorrelation can arise even if weaker
absorbers are not associated with galaxies in any way.

Taken with the above discussion of the selection biases 
inherent in the L95 and C98 distributions of Figure~\ref{Lanzetta}, 
we conclude that, while some of the highest column density \lya 
systems in the local Universe may be associated with the gaseous 
halos of nearby bright galaxies, the lower column density clouds 
show no evidence of extending this correlation. 
In fact, \citet{Dave99} have reproduced the salient features of 
Figure~\ref{Lanzetta}, including the observed slope shown by the dotted line,
 using a N-body plus hydrodynamical simulation 
in which mass feedback from galaxies into the IGM is small. 
In other words, both the L95/C98 correlation at 
high \W and the increasing scatter at low \W in our data can be
explained by these simulations. In \citet{Dave99} this correlation is
reproduced simply because galaxies and \lya clouds both trace the 
same large-scale structure. The simulations also predict that the
correlation should degrade towards lower column density because three
distinct populations of clouds, with somewhat different relationships
to galaxies, are involved: (1) cold, condensed clouds that are quite
close to galaxies ($\Dperp\leq 30\hsfi $kpc) and are plausibly galaxy 
halos; (2) highly-ionized absorbers, shock-heated by infalling
large-scale structure and thus near ``filaments" of galaxies
($30\hsfi \leq\Dperp\leq300\hsfi $kpc); and (3) warm, diffuse photoionized clouds
that are truly intergalactic and thus much more randomly placed relative to
galaxies ($\Dperp\geq300\hsfi $kpc). The L95 and C98 absorber samples
appear to contain primarily type 1 and 2 clouds, while the GHRS studies
consist of types 2 and 3. See IPF99 for a speculative division of local
\lya absorbers into these three types described by \citet{Dave99}. 

The C98 results that the correlation  in Figure~\ref{Lanzetta} are largely 
independent of the properties  of the nearby galaxy lends further support for the contention
that  this correlation is the result of large-scale structure alone, as
\citet{Dave99} predict. 
C98 found only a correlation with galaxy luminosity, but not with any
star formation properties (e.g., Hubble type or color), which would certainly be expected 
if these absorbers were due to some pervasive feedback of gas from galaxies
into the IGM. Additionally, the ``failure'' of the sensitive 
galaxy searches around \lowz \lya absorbers by 
\citet{Rauch96}, \citet{VG93,VG96}, \citet{Shull96,pks}, and IPF99 plus the
examples of local \lya absorbers found in galaxy voids by \citet{Stocke95}
and this paper, adds considerable evidence to the contention that not all \lya clouds 
arise in extended gaseous galaxy halos.

Finally, while some authors have suggested that many low column absorbers
might be associated with low surface brightness (LSB) galaxies \citep{Impey97b,Linder00},
there are no current data supporting this suggestion. Specifically,
\citet{Rauch96} searched the 3C~273 field for LSB galaxies at Virgo distances and found none;
IPF99 found similar negative results in the Virgo cluster proper. 
Since many LSB galaxies are \hone rich, they should be detected easily by
\hone images of \lya absorber locations and have not been \citep{VG93,VG96,Shull96,Hibbard02}.
New \hone and optical surveys are now in progress to address this suggestion further.
\clearpage
\section{\lya absorbers and Galaxy Filaments} \label{sec:filaments}
\subsection{The 3C~273/Q~1230+0115 field and absorbers}\label{sec:3Q}
The proximity of the 3C~273 and Q~1230+0115 sightlines, 0.91\degr\ apart 
in the sky, provides an opportunity to look for common 
absorbers at very low \z\ and thus to infer the sizes of \lya 
clouds. Due to the depth of galaxy surveys in this region, these two 
adjacent sightlines also allow an important new method for relating \lya
absorbers to galaxies and large-scale structure. 
Over the range 
$1000 < cz < 2300$\kmsno, these sightlines are separated by
(230--550) \hsfi kpc. Absorbers in that redshift range are located in
the southern portion of the Virgo Cluster region. 
Updated column densities for the two strongest Virgo cluster absorbers
along the 3C~273 sightline have been reported by \citet{Sembach01} on the basis
of FUSE spectra.  

\begin{figure}[htb]
  \plotone{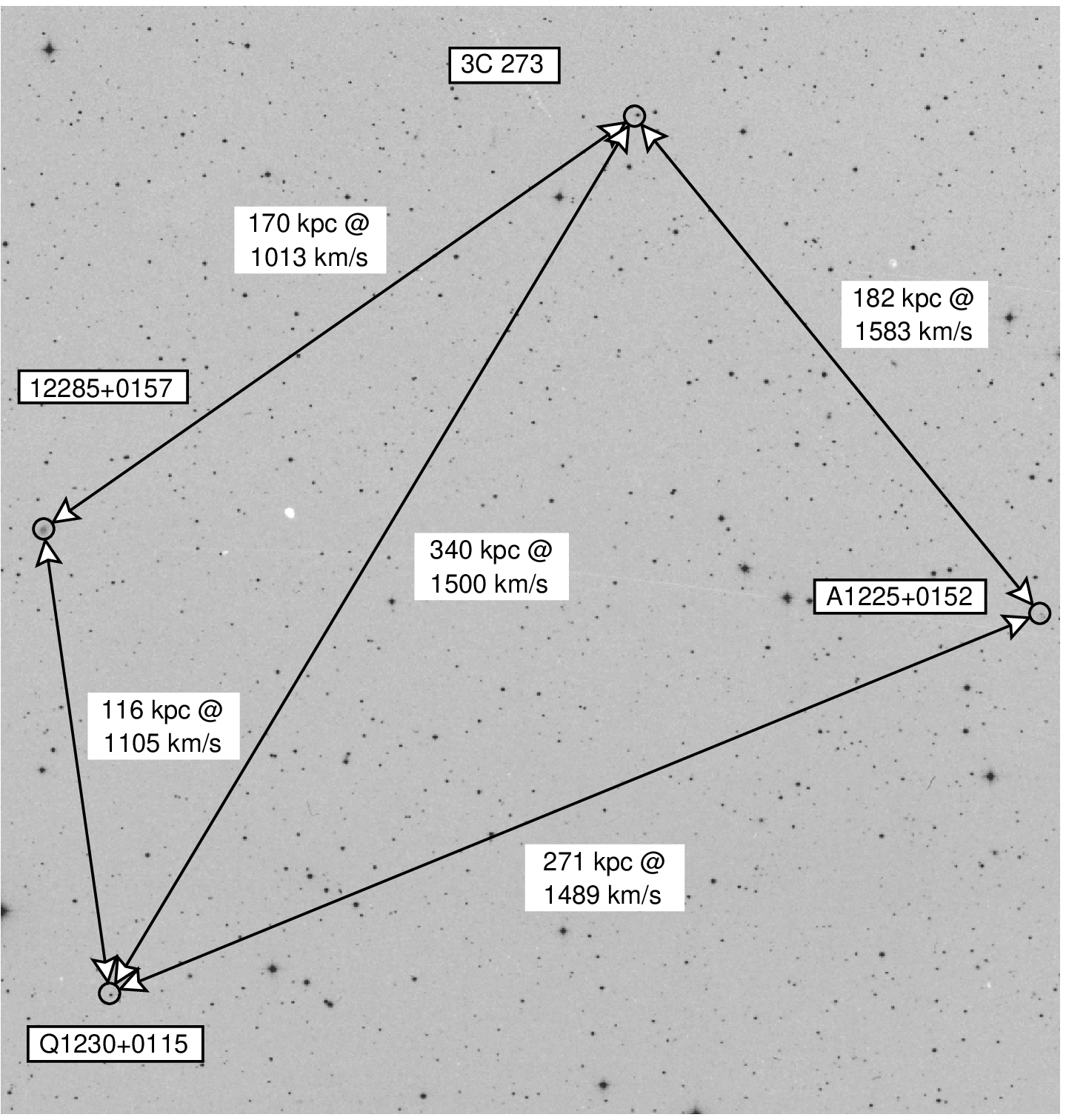}
  \caption{\label{3CQ} Digitized Sky Survey image of the 
3C~273/Q~1230+0115 field. The image is $0.9\degr\ \times 1.0\degr$. 
Perpendicular distances (\Dperp) between the sightlines and
from the A1225+0152 and 12285+0157 galaxies to the sightlines are indicated
for \Hnaught=70 \hseventy\kmsno\perMpcno.}
\end{figure}

\begin{figure}[htb]
  \plotone{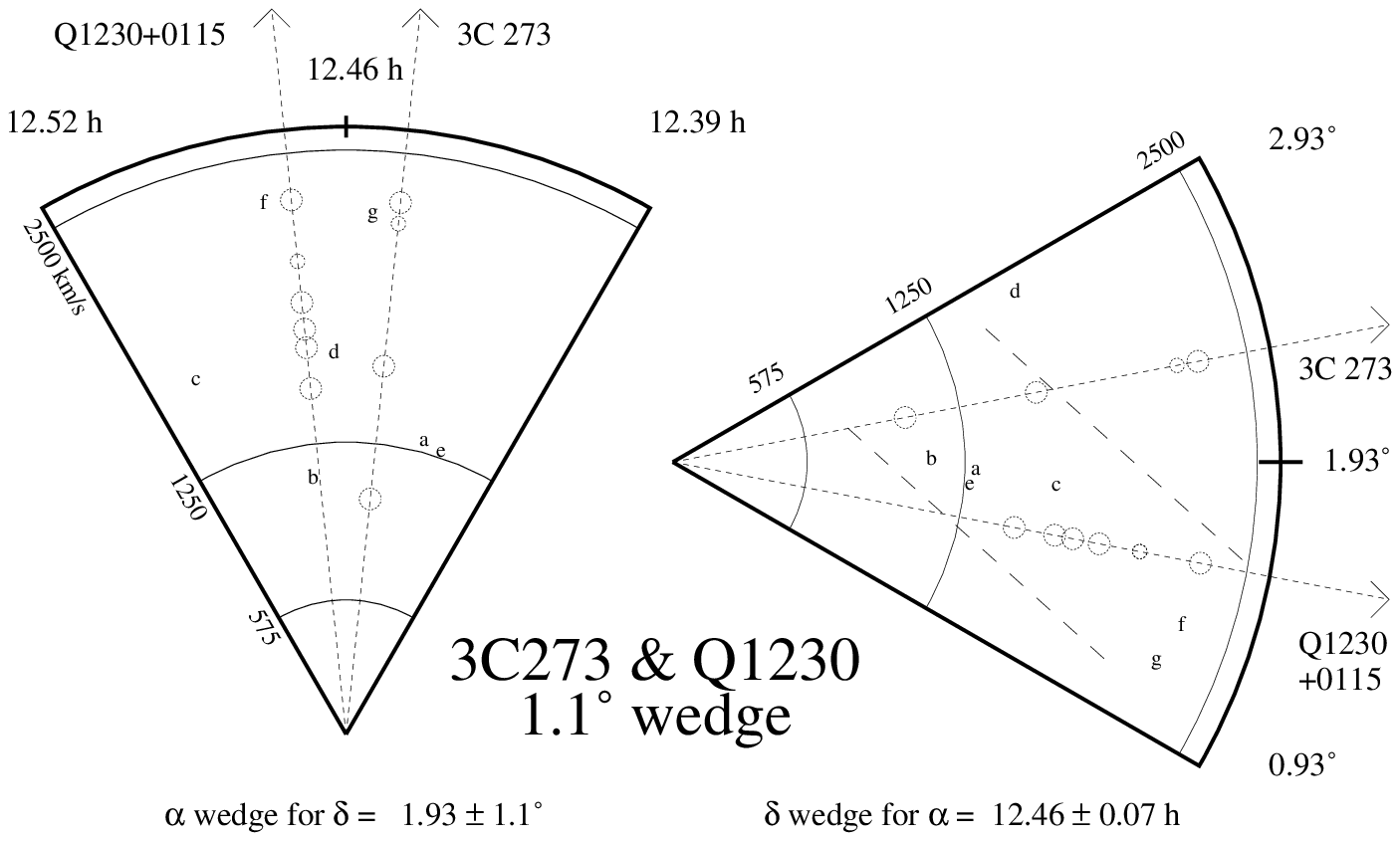}
  \caption{\label{3CQ_pie}  1.1\degr\ pie diagram showing the 3C~273 
and Q~1230+0115 sightlines, nearest galaxies, and absorbers with 
\vhelio \lt 2600\kmsno. The \lya absorbers are located with dashed
circles; the larger circles are \foursig detections, the smaller circles
are \threesig detections. The arc at 575\kms indicates the lower velocity limit 
of our \lya detections, corresponding to our spectral limit of 
$\lambda \ge 1218\ang$. The long-dashes in the declination 
(right) pie diagram outline the possible filamentary structure associated with both galaxies and absorbers.}
\end{figure}

A digitized sky survey (DSS) image around 3C~273 
and Q~1230+0115 is presented in Figure~\ref{3CQ}.
Additionally, 1.1\degr\ ``pie diagrams" for this sightline, centered halfway between the 
two targets, are shown in Figure~\ref{3CQ_pie}. Unlike the pie 
diagrams in Paper~I, each galaxy is given its own designating letter in 
Figure~\ref{3CQ_pie}. Table~\ref{3CQfield}
lists these galaxies and their relationship to the 3C~273 and 
Q~1230+0115 sightlines, including by column: 
(1) the galaxy name in the CfA redshift catalog, 
(2) the galaxy's heliocentric radial velocity in\kmsno;
(3) and (4) the distance in \hsfi kpc on the sky between the galaxy 
and the 3C~273 and Q~1230+0115 sightlines, respectively; 
(5) the letter designation for the galaxy as shown in Figure~\ref{3CQ_pie};
(6) and (7) the heliocentric velocity of the closest \lya absorber in the 3C~273 and
Q~1230+0115 sightline respectively. 
Galaxies (a) and (b) are the
closest to these sightlines and are shown in the image in Figure~\ref{3CQ}, as
well. Galaxy (a),
A1225+1052, is also known as the Haynes \& Giovanelli \hone cloud
\citep{Giovanelli91} and is composed of two \hone peaks, only one of
which clearly contains starlight \citep{Salzer91}. Galaxy (b),
12285+0157, has been recently studied by \citet{Hoffman98}. 
Because of infall to the Virgo Cluster, velocity-dependent distance estimates of 
galaxies in this region are uncertain on the order of \about 1-2\hsfi Mpc; 
we assume that these ``streaming motions" affect galaxies and \lya
absorbers the same and do not introduce additional errors to the galaxy-absorber distances. 
In this Section we will continue to estimate distances
using recession velocities and $\Hnaught=70\kmsno\perMpcno$.  

A 1\degr\ area around the 3C~273 sightline was surveyed for galaxies 
by \citet{Morris93} complete to $m_B\sim19$. This region lies within
the southern extension of the CfA survey made by \citet{GGH98}. 
At $20\hsfi$ Mpc or \cz =1400\kmsno, the \citet{Morris93} magnitude
limit implies that galaxies have been surveyed in 
this field down to $\sim10^{-3}$ \Lstarno.  
Further redshift survey work towards the 3C~273 and Q~1230+0115 sightlines is in progress
\citep{McLin02}. 
\begin{deluxetable}{lcccccc}
\tablecaption{Galaxies in the 3C~273/Q1230+0115 Field\label{3CQfield}}
\tabletypesize{\footnotesize}
\tablecolumns{7}
\tablewidth{0pt}
\tablehead{
\colhead{Galaxy Name} & 
\colhead{\cz$_{gal}$\tablenotemark{a}} & 
\colhead{\Dperp\tablenotemark{b}~~~~to} & 
\colhead{\Dperp\tablenotemark{b}~~~~to} & 
\colhead{Label\tablenotemark{c}} &
\colhead{\cz$_{abs}$\tablenotemark{d}} & 
\colhead{\cz$_{abs}$\tablenotemark{e}} \\
\colhead{}&
\colhead{(\kmsno)}&
\colhead{3C~273} &
\colhead{Q1230+0115} &
\colhead{} &
\colhead{3C~273} &
\colhead{Q1230+0115}
}\startdata
A1225+0152  & 1298 & 182   &271   &a & 1012 & 1487 \\
12285+0157  & 1105 & 170		 &116 &b & 1012 & 1487 \\
12308+0147  & 1643 & 483   &279  &c & 1583 & 1663 \\
12277+0254  & 1635 & 260   &562  &d & 1583 & 1663 \\
A1224+0146  & 1275 & 282   &294  &e & 1012 & 1487\\
A1228+0116  & 2289 & 678	  &162  &f & 2287 & 2298 \\
12270+0107  & 2234 & 677   &293  &g & 2287 & 2298 \enddata
\tablenotetext{a}{Galaxy recession velocity.}
\tablenotetext{b}{Galaxy to line-of-sight (LOS) distance, perpendicular to the LOS (in \hsfi\ kpc).}
\tablenotetext{c}{Label corresponding to Figure~\ref{3CQ_pie}.}
\tablenotetext{d}{Velocity of nearest \lya absorber along the 3C~273
sightline}
\tablenotetext{e}{Velocity of nearest \lya absorber along the Q~1230+0115
sightline}
\end{deluxetable}

All eight definite absorbers (3 towards 3C~273 and 5 towards Q~1230+0115) and two
possible \lya absorbers (one in each sightline) are presented in the ``pie diagram'' of 
Figure~\ref{3CQ_pie} as large and small dashed circles respectively. 
Considering only the definite absorbers, there are two close velocity 
matches between the sightlines: 
the 1583\kms (3C~273) absorber matches both with the 1489 and 1665\kms 
(Q~1230+0115) absorbers. Additionally, a possible absorber at 2287\kms 
(3C~273) is closely matched with a definite absorber at 
2300\kms (Q~1230+0115), which needs better 3C~273 data to confirm.
At lower spectral resolution (e.g., FOS at 1\ang), the pair of close absorbers (1489
and 1665\kmsno) in the Q~1230+0115 sightline would be blended and so would be paired with the 1583\kms absorber in 
the 3C~273 sightline to form  
a single pair of ``common absorbers''. This pairing would 
suggest a transverse size for this \lya cloud of
$\sim350\hsfi $kpc, similar to values found at higher redshift 
\citep[e.g.,][]{Dinshaw95,Impey97}.
However, four definite absorbers
have no counterpart in the other sightline, suggesting upper limits on
\lya absorber sizes comparable to, or smaller than, this  value.  
 
The information in Figure~\ref{3CQ}  can be interpreted in terms of the
extended galaxy halo hypothesis of L95, in that both galaxies (a) and
(b) are close enough to both sightlines to expect \lya absorbers in both
spectra by the hypothesis of L95. Indeed, one could plausibly assign both the
1489\kms or 1665\kms absorbers in Q~1230+0115 and the 1012\kms or 1583\kms absorbers in
3C~273 to A1225+0152 ($\cz_{gal} = 1298 \pm 20 \kmsno$), giving that galaxy a
$\sim300\hsfi $kpc gaseous halo down to approximate column densities of 
$10^{14}\percmtwono$. Although three of the four absorptions are at higher recession
velocities than the systemic velocity of A1225+0152, 
the fact that this galaxy is actually two
interacting systems is supporting evidence for this picture.
The second \hone emission peak is located $\sim100\hsfi $kpc to the
SW of A1225+0152 \citep{Giovanelli91}, and thus just off the right edge of Figure~\ref{3CQ}
with an elongated \hone halo of diameter 215\hsfi kpc at $\Nhno = 2 \times 10^{18}$\percmtwono. 
It has been speculated that galaxy interactions can
increase greatly the gaseous cross-sections of galaxies
\citep{Mv,VG96}, and this 
set of galaxies and absorbers offers a laboratory to test that hypothesis. 

However, the other nearby galaxy, 12285+0157, offers a counter-example
to this same hypothesis. While the 3C~273 spectrum has an absorber which
plausibly could be associated with this galaxy, being only 100\kms lower
than the galaxy velocity, the spectrum of Q~1230+0115 has no absorption
within several hundred\kms of the galaxy redshift. This is despite the
fact that A12285+0157 is closer to the Q~1230+0115 sightline than to that
of 3C~273 by a factor of 1.5. At the very least, these two sets of absorbers and galaxies
suggest that very extended galaxy halos (if they exist) cannot be spherical with unity
covering factor. This is contrary to the suggestion by L95 and C98 that
galaxy halos are spherical with large covering factor (see \S~\ref{sec:stat_filaments} 
and \citet{Bowen96} for further discussions).

Figure~\ref{3CQ_pie} suggests an alternative explanation, described by the
dashed lines, in which galaxies and clouds in this region occupy a single
large ``filament'' $\gt20\hsfi $Mpc long, with an aspect ratio of \gt20:1.
This possible alternative interpretation makes all but one of the observed definite
absorbers ``common'' between these two sightlines. A further statistical
test that \lya absorbers lie in galaxy ``filaments'' will be presented in
the next section.
We await better S/N and resolution spectra for both of these targets,
as well as more galaxy redshift survey data in this region to confirm
this possible ``filament".
Figure~\ref{3CQ_pie} shows the value of performing sightline pair
experiments at low \z, where both galaxies and absorbers can be 
integrated into a common picture of large-scale structure.

\subsection{Statistical Tests of \lya Cloud Alignment with Galaxy Filaments}
\label{sec:stat_filaments}
In this section we investigate the statistical relationship between
local \lya absorbers and large-scale structure using the distribution of 
CfA galaxies in (3-11) \hsfi Mpc diameter spherical or
cylindrical regions around each absorber. The purpose of this investigation is
to determine whether the local \lya absorbers preferentially align with
the large-scale ``filaments'' of galaxies,
to which the \lya clouds are also plausibly related in numerical simulations
\citep{Dave99}. A possible example of such an alignment was
presented in the last subsection. 

Because the methodology used in this section is new and somewhat complicated,
an overview is supplied in this and the next paragraph. The basic approach is similar 
to comparing the position angle of the absorber to galaxy impact parameter vector 
(``absorber PA'') to the galaxy major axis, as in C98. Instead we use many 
nearby galaxies in a specified volume to compare the mean ``absorber PA'' to  the 
major axis of the galaxy large-scale structure ``filament'', composed of these same galaxies. 
Much of the below discussion is devoted to describing how the
``filament'' major axis is determined using the mean PA of all galaxy-galaxy 
vectors. We do not presume that filamentary structure is present in each volume
beforehand, just as measurement of a galaxy major axis does not presume a highly flattened structure. 
We simply use whatever galaxy distribution is present to find the ``filament major axis'' using the 
mean of all galaxy-galaxy vectors. As described in detail below, this procedure  returns a 
distribution of PAs which, in the mean, has a full-width-half-maximum of $\pm 20\degr$. 
An important variable in this process is the size of the volume around each absorber, which
determines the number of galaxies (N) to be included in this analysis. For our fiducial 
volume (4.3\hsfi Mpc scale size), a typical absorber has several tens of galaxies in
this volume but ``void absorbers'' have only a few, while some very nearby absorbers in
very well-surveyed galaxy regions have several hundred (we truncate $N\leq$200 to avoid
any few absorbers from dominating our statistics). Through experimentation we have found
that volumes $\leq3\hsfi$ Mpc in scale size contain too few galaxies to determine an 
accurate ``filament'' PA; volumes $\geq8\hsfi$ Mpc in scale size are so big that they 
contain several ``filaments'' (this is evident in ``pie diagrams'' from the CfA redshift
survey shown in Paper 1), also degrading the accuracy of the ``filament'' PA. For volumes
between these two extremes the ``filament'' PA can be determined to a mean accuracy of 
$\pm$5 degrees (see discussion below).

Once the ``filament'' and ``absorber'' PAs are determined for each volume, we align all
vectors to the ``filament'' PA and co-add the vectors from  all volumes in the \sample 
absorber sample. We then cross-correlate the ``filament'' PAs with
the ``absorber'' PAs. We then repeat this procedure using ``absorber'' PAs for a set of
``fake'' absorbers, randomly located in the same volumes, to supply a comparison 
baseline. The amount by which the cross-correlation function (CCF) for the real absorbers 
exceeds the CCF for the ``fake'', randomly-distributed, absorbers is the statistical
significance to which absorbers are aligned with galaxy ``filaments''. 
Large statistical significances are found for most volumes, as described in detail in
the paragraphs below.

To explore the possible relationship between the detected \lya 
absorptions and galaxy filaments, 
we compared our \sample sample of absorbers to the 
large-scale galaxy distribution in the vicinity of each absorber.
To do this, we extracted a constant-volume cylinder or sphere 
from the CfA survey centered on each 
absorber.  The diameter of the sphere and the length and diameter of 
the cylinder were set to \Dv=300 or 450\kms (4.3 or 6.4\hsfi Mpc). 
The choice of \Dv=300\kms matches our previous ``retarded Hubble flow'' 
model of $\pm$300\kms and appears to be a good match to the scale of large-scale
structure ``filaments'' (see below). 
Additional larger box sizes were also examined 
to correct for the uncertainties associated with
inferring galaxy/absorber distances from redshifts without knowledge 
of their peculiar velocities (PVs). This is done by allowing an 
additional $\pm$150\kms or $\pm$300\kms in
PV for both our spherical and cylindrical 
models. The rationale behind these modified models is to include filament
galaxies with large peculiar velocities, which would, if interpreted as pure Hubble flow velocities, 
place them just outside the sample volume. But, this also includes many more
galaxies with low PV at larger distances from the absorbers;
i.e., with increasing PV allowance, the region of space and number of galaxies included 
in our sample increases. 
In addition, we explored the consequences of different 
extraction volumes by performing trials for smaller (\Dv=200\kmsno) and 
larger (\Dv=600 and 800\kmsno) sizes. We find that volumes smaller than 300\kms
are too small to include a large enough number of
galaxies to define adequately the large-scale structure, 
while larger volumes ($\ge$ 600\kmsno) include more than one large-scale structure ``filament'', making
the large-scale structure more complicated to interpret (see the ``pie diagrams''
in Paper~I).

For all CfA galaxies in the extracted volume, we first determine
the best-fitting ($\alpha,\delta,\cz)$ plane ($\cz = A+B\alpha+C\delta$)
of the galaxies in the volume using a robust, \Xtwo-minimizing fit. 
The algorithm reduces the impact of ``outliers'' which do 
not reside near the dominant plane, if any, of the galaxies in each extracted volume.
Before determining the best-fit plane, we convert $\alpha, \delta$ and \cz\ 
to units of  \hsfi Mpc.
Once the dominant plane has been 
identified, we rotate the galaxy coordinates (X,Y,Z) in Mpc so that 
the dominant plane is aligned perpendicular to the new Z$^{'}$ axis for viewing.
This alignment to the coordinate system (X$^{'}$,Y$^{'}$,Z$^{'}$) 
attempts to remove any viewing angle effect and reduces our 
exploration to the simple question: When viewed
from a consistent ``face-on'' viewing angle, do our detected absorbers show any 
preference to be aligned with the filamentary structure seen in
galaxy clustering?

Once the extracted CfA regions have been rectified to consistent, 
face-on viewing angles, we calculate 
the two-dimensional (2D) position angles (PAs)  
for all galaxy-galaxy and absorber-galaxy pairs to produce  
the distribution of PAs for each volume.
For all absorber volumes, the dominant galaxy-galaxy PA is
located by fitting a Gaussian to the PA distribution and choosing the Gaussian center as the
mean PA of the filament. 
The ``raw'' PAs are then adjusted to the Gaussian mean PA 
so that the adjusted distribution is 
centered at PA=0\degr; i.e., these adjusted, or aligned, PAs (APAs) are relative to the major axis of
the dominant galaxy filament in each volume. 
Because all absorber volumes now have their PAs zero-pointed to the best-fit, local
filament PA, we can co-add the galaxy-galaxy and absorber-galaxy APAs from all volumes into 
a single distribution. To prevent any one absorber volume 
from dominating the APA distribution, we use a randomly selected subsample of only 200
galaxies for any volume with more than 200 galaxies. 
The median number of galaxies in each region for each model are given in Table~\ref{CCFRES}.
 
\begin{figure}[htbp]
  \plotone{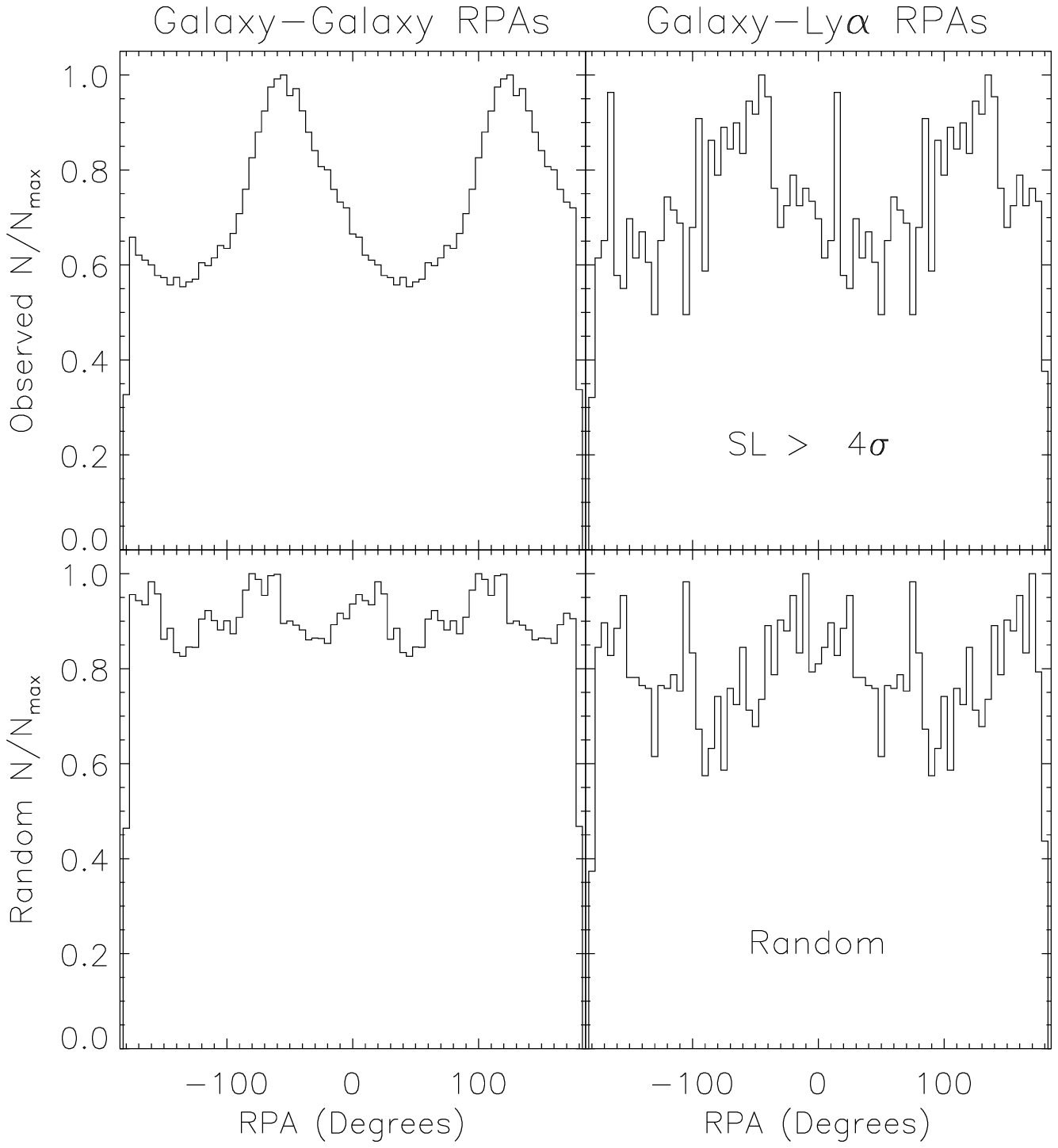}
  \caption{\label{RAW_PA} ``Raw'' galaxy-galaxy (left column) and 
absorber-galaxy (right column) position angles (PAs) 
using the spherical volume sampling described in the text, with \Dv=450\kms 
and an additional $\pm$150\kms peculiar velocity. These PAs are ``raw'' in
the sense that they have not been zero-pointed to the dominant PA of the
galaxy distribution in each volume.
The upper two panels correspond to our \real\ \lya sample, while
the lower panels refer to a random distribution. The large feature 
in the ``raw'' distributions at -55\degr\ and +125\degr\ is 
the ``Great Wall'', through which many of our sightlines pass.}
\end{figure}

\begin{figure}[htbp]
  \plotone{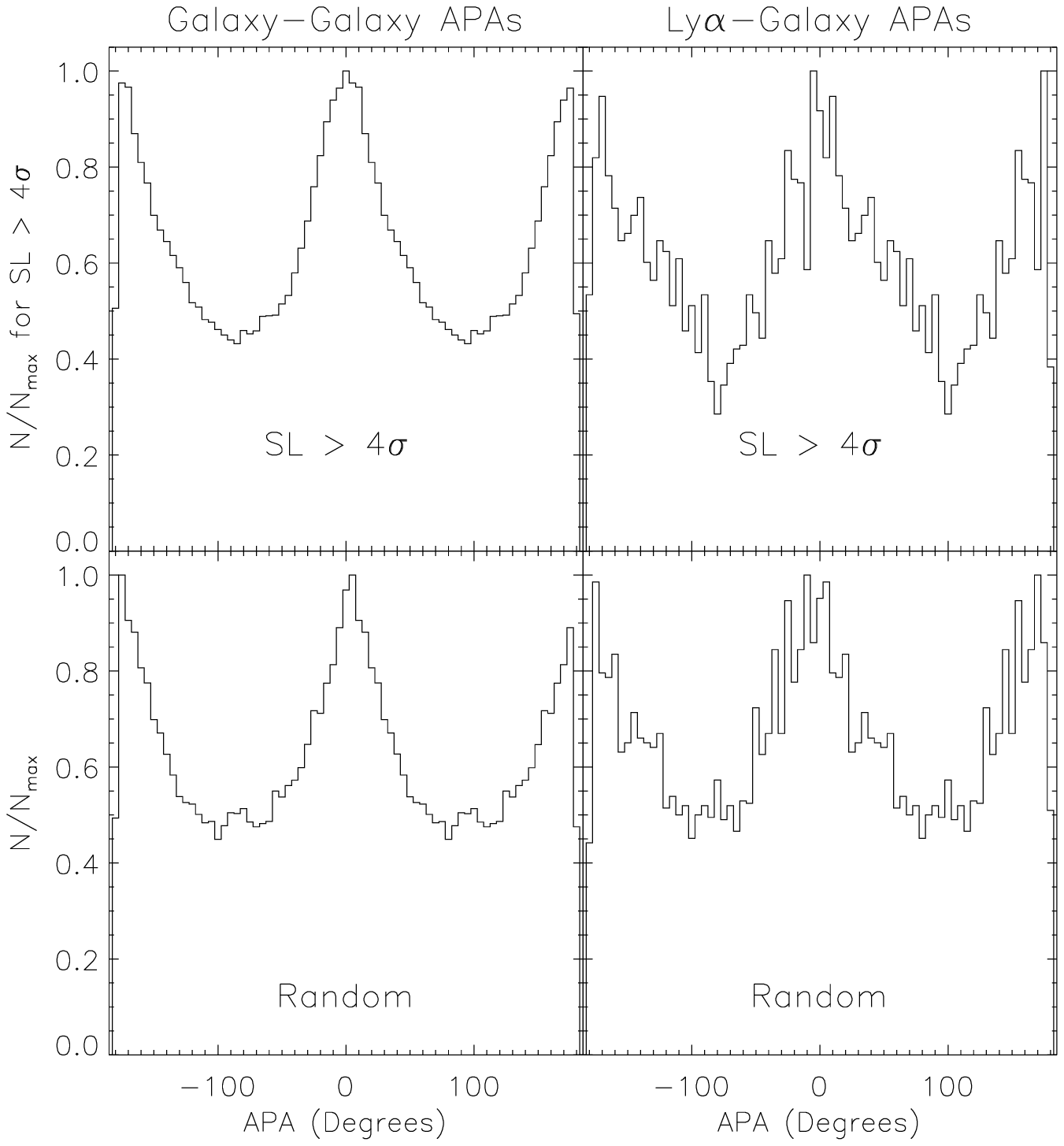}
  \caption{\label{ADJ_PA} Aligned galaxy-galaxy (left column) and 
absorber-galaxy (right column) position angles (APAs) using 
the spherical volume sampling described in the text, with \Dv=450\kms and  
PV = $\pm$150\kms peculiar velocity. The upper two panels correspond 
to our \real\ \lya sample, while the lower panels refer to a simulated random 
distribution of absorber locations. 
Since these PAs have been adjusted to a zero point corresponding
to the dominant PA of galaxy-galaxy vectors in each volume, the large peaks
in these distributions indicate the same strong filamentary structure 
in the galaxy distribution along our sightlines.}
\end{figure}

Figure \ref{RAW_PA} shows the ``raw'' PAs (RPA) for both the
galaxy-galaxy and absorber-galaxy pairs in a spherical volume
model with PV = $\pm$150\kms and \Dv = 450\kms diameter. 
Figure \ref{ADJ_PA} presents the APAs for 
the same model, and shows that regardless of absorber location, the galaxy-galaxy PAs
have a very narrow peak ($\pm20\degr$ FWHM). Thus, in the mean, the ``filament'' PA can be
determined with an accuracy that is a fraction of this; i.e., $\pm5\degr$ conservatively.
The rather narrow peak for the galaxy-galaxy APA distribution in Figure~\ref{ADJ_PA} tells
us that the filaments in our sample are not predominantly ``sheetlike''; i.e. they do not
fill the viewing plane when rectified to ``face on''.
The broader distribution in the \lya-galaxy APAs is simply due to the inherent number statistics
(i.e., N versus ~N$^2$/2).
Also shown in Figures~\ref{RAW_PA} and \ref{ADJ_PA} 
are ``raw'' PAs (RPAs) and APAs, respectively, for a comparison sample of random
absorber locations within volumes of the same size and shape as the
volumes from which the observed PA distributions were extracted.   
These random locations along our actual GHRS sightlines
were constructed in an identical way as in 
Paper~II and \S~\ref{sec:relationship}, accounting for the varying
sensitivity and spectral coverage of our observations. 
Random APAs were also constructed for sightlines randomly placed throughout
CfA survey regions of sky, using the median sensitivity 
function of our observations. This approach approximately matches our HST/GHRS  
observations in terms of number 
of sightlines passing through well-surveyed CfA regions (see \S~\ref{sec:sample})
from $575 \le cz \le 10,000\kmsno$. The random APAs from
these ``all-sky" sightline simulations are nearly identical to our random
individual sightline results.  


To determine the significances of the differences between the 
galaxy-random absorber, galaxy-real absorber, and galaxy-galaxy APA distributions, 
we compare the cross-correlation functions (CCFs) of the APAs. For the
 galaxy-galaxy APAs, this is obviously the auto-correlation function.
For each of our random methods, using our sightlines and the random (``all-sky'') method, we performed
N$_{trial}$ Monte Carlo simulations, 
each one simulating our combined HST/GHRS observations, using the median 
sensitivity function, \sensno, and our \dndW\ results from Paper~II.
The value of N$_{trial}$ for each model is given in Table~\ref{CCFRES}. For the
``all-sky" trials, the median \sens was applied. 

\begin{figure}[htbp]
  \plotone{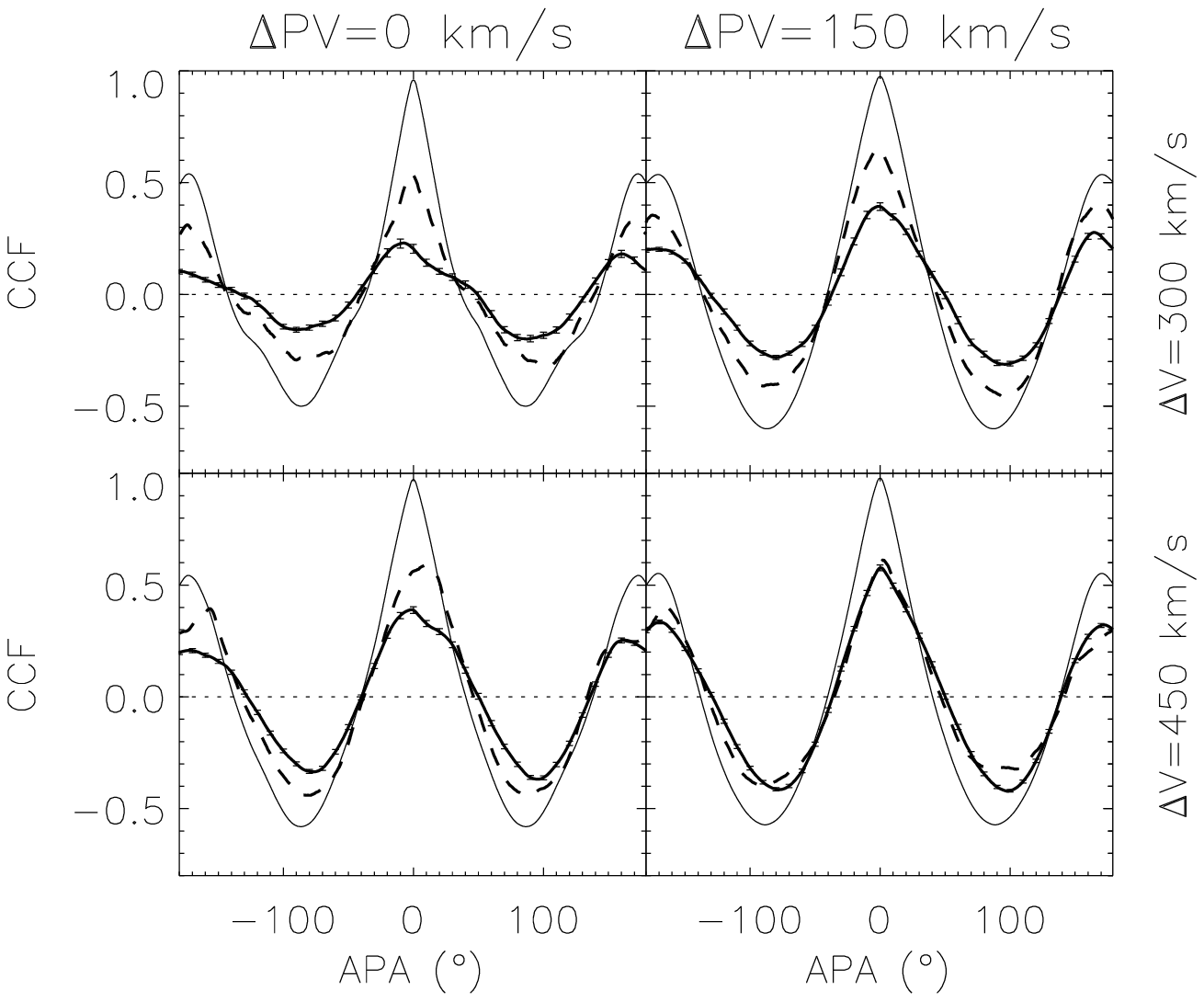}
  \caption{\label{CCFcompare} CCFs of the adjusted galaxy-galaxy (line solid line), absorber-galaxy (dashed line) and 
random absorber-galaxy (heavy solid line with \onesig error bars) position angles (APAs) using 
the spherical volume sampling described in the text. The upper panels show the
CCFs for \Dv=300\kms (4.3\hsfi Mpc) and 0\kms (left) and $\pm$150\kms (right) peculiar velocity (PV) allowances. 
The bottom panels correspond to \Dv=450\kms (6.4\hsfi Mpc). The random absorbers in this figure were 
constructed using our ``all-sky" method (see text).  The detected absorber significance levels 
 above the  random distributions for each trial are given in Table~\ref{CCFRES}.}
\end{figure}

In Figure~\ref{CCFcompare} we present the CCFs for the \Dv=300 and 450\kmsno, 
PV=$\pm$0 and $\pm$150\kms spherical models. 
Since these APAs have been adjusted to a zero point corresponding
to the dominant PA of galaxy-galaxy vectors in each volume, the large peaks
in these distributions indicate the same strong filamentary structure 
in the galaxy distribution along our sightlines as seen in Figure~\ref{ADJ_PA}.
To calculate the significance of
our galaxy-absorber CCF over the random distribution, we compare the area in 
the central CCF peak to the random distribution. 
In Figure~\ref{CCFcompare}, the (small) error bars represent the 1\sig standard deviations 
of the random ``all-sky" CCFs. The extent of the central peak,
over which we calculate the excess power in the absorber-galaxy CCF,
is defined by the locations of zero CCF power in the galaxy-galaxy auto-correlation function
(the filament width as defined by the galaxy locations).
By dividing the difference in areas by the \onesig uncertainty inferred from
the random trials, we determine the degree to which absorbers (and galaxies) align
with filamentary large-scale structures. These absorber results are given in
Table~\ref{CCFRES}. Particularly, for the smaller volumes with no PV allowance,
the excess power is large; $5-12\sigma$ for the various models.
The PV = $\pm$300\kmsno, \Dv=450\kms model creates a sample region 
that is the equivalent of 1500\kmsno, or 21.4 \hsfi Mpc along the line of sight. 
As seen from the pie diagrams in Paper~I, this volume is much
larger than most filamentary structures, hampering our ability to correctly
identify the dominant filamentary plane and the dominant PA necessary 
for proper PA zero-pointing. 
Thus, we believe that the \Dv=450\kmsno, PV $= \pm 300\kmsno$, and larger volume models,
sample too large a volume to   determine accurately the local filamentary structure, 
causing our random and absorber  results to appear similiar for both the spherical and cylindrical models.
On the other hand, for the \Dv$\lt$200\kms volume with no PV, 
there is little power in the random absorber CCF distribution; i.e., this volume is too small to allow the PA of the
large-scale filamentary structure in the CfA catalog to be determined accurately. Thus, we believe that 
the first three models of each type (cylindrical, or spherical volume) in Table~\ref{CCFRES} give the most
accurate representation of absorber-galaxy filament alignment.
\begin{deluxetable}{ccc|crc|crc}
\tablecaption{\label{CCFRES}Cross-Correlation Results for our \real\ Sample}
\tabletypesize{\scriptsize}
\tablewidth{0pt}
\tablecolumns{9}
\tablehead{
\colhead{\Dv} & 
\colhead{PV\tablenotemark{a}} & 
\colhead{Random} & 
\multicolumn{3}{|c|}{Spherical Model} &  
\multicolumn{3}{c}{Cylindrical Model} \\
\colhead{(\kmsno)}& 
\colhead{(\kmsno)}& 
\colhead{Method\tablenotemark{b}}& 
\multicolumn{1}{|c}{SL\tablenotemark{c}} & 
\colhead{N$_{trials}$\tablenotemark{d}} & 
\colhead{median(N$_{gal}$)} & 
\multicolumn{1}{|c}{SL} & 
\colhead{N$_{trials}$} & 
\colhead{median(N$_{gal}$)}}
\startdata
300&  0&A& 12.6& 41&  74 &  14.8& 44&  74 \\
300&150&A& 11.1& 16&  76 &  6.8& 15&  76 \\
450&  0&A& 12.0& 19&  72 &  8.3& 27&  70 \\
450&150&A& $<$ 1.0& 29&  74 & $<$ 1.0& 33&  73 \\
\tableline
300&  0&G&  8.4& 41&  68 & 11.8& 44&  68 \\
300&150&G& 10.2& 16&  69 &  6.0& 15&  69 \\
450&  0&G&  9.1& 19&  68 &  8.3& 27&  69 \\
450&150&G& $<$ 1.0& 29&  70 & $<$ 1.0& 33& 71 \enddata
\tablenotetext{a}{Peculiar velocity allowance.}
\tablenotetext{b}{Method used to generate the random absorber sightlines; G (GHRS) uses our GHRS sightlines; 
A (``all-sky") uses a random distribution of sightlines through
well-sampled regions of the CfA catalog.}
\tablenotetext{c}{Significance Level above random as defined in the text.}
\tablenotetext{d}{N$_{trials}$ is the number of random trials used. Each random trial matches the number 
and pathlength of our GHRS sightlines.}
\tablenotetext{e}{N$_{gal}$ is the number of galaxies per detected absorber.}
\end{deluxetable}

These results agree with results from a simpler two-dimensional analysis 
which only considers position angle on the sky, without compensation for
recession velocity. As seen from the significance levels in Table~\ref{CCFRES}, these
trials provide evidence that  \lya absorbers align with galaxy filamentary structures in the local Universe. 
Together with the lack of any correlation between low column density local \lya absorbers
and individual galaxies, this
result is further proof that the \lya absorbers are related to large-scale structure,
not to individual galaxies, in agreement with numerical simulations. 
This result should also be contrasted with the C98 result that there is no statistical 
relationship between the ``absorber PA'' and the major axis of the nearest, bright
galaxy. C98 interpret their result as meaning that galaxy halos are spherical with near
unity covering factor. But, using a different technique, \citet{Bowen96} find a 
much lower covering factor of 0.4. Therefore,  in the light of the present results,
where an alignment between \lya absorbers and galaxy filamentary structure has been found,
the random location of \lya absorbers with respect to the nearest galaxy may not be indicative 
of a galaxy halo property at all.
\clearpage
\section{\dndz of \lya absorbers in Voids}\label{sec:dndz_voids}
In this section, we explore the line-of-sight frequency \dndz as a function of \W
 for the \lowz \lya forest divided into void and ``supercluster'' subsamples.  
As before, we use only our \sample sample and define a void location as a region of space with no known galaxy within
3\hsfi Mpc. 
We have explored using other methods to define voids along our sightlines, guided by the
\citet{Slezak93} wavelet transform method for defining voids in the original CfA survey.
Since several of our GHRS sightlines pass through this region, we have used the 
\citet{Slezak93} void boundaries to determine that a $\pm 300\kms$ allowance from the
edges of the supercluster filaments as defined by galaxy locations is an 
adequate representation of the void boundaries along our sightlines.
Alternatively, $\pm 1000\kms$ from the filament center is a slightly more
conservative definition of the void boundaries
(it yields slightly smaller voids than the first two definitions).
The results stated below are for the simple 3\hsfi Mpc definition, 
but all three definitions give the same result within the errors.

\begin{figure}[htbp]
  \plotone{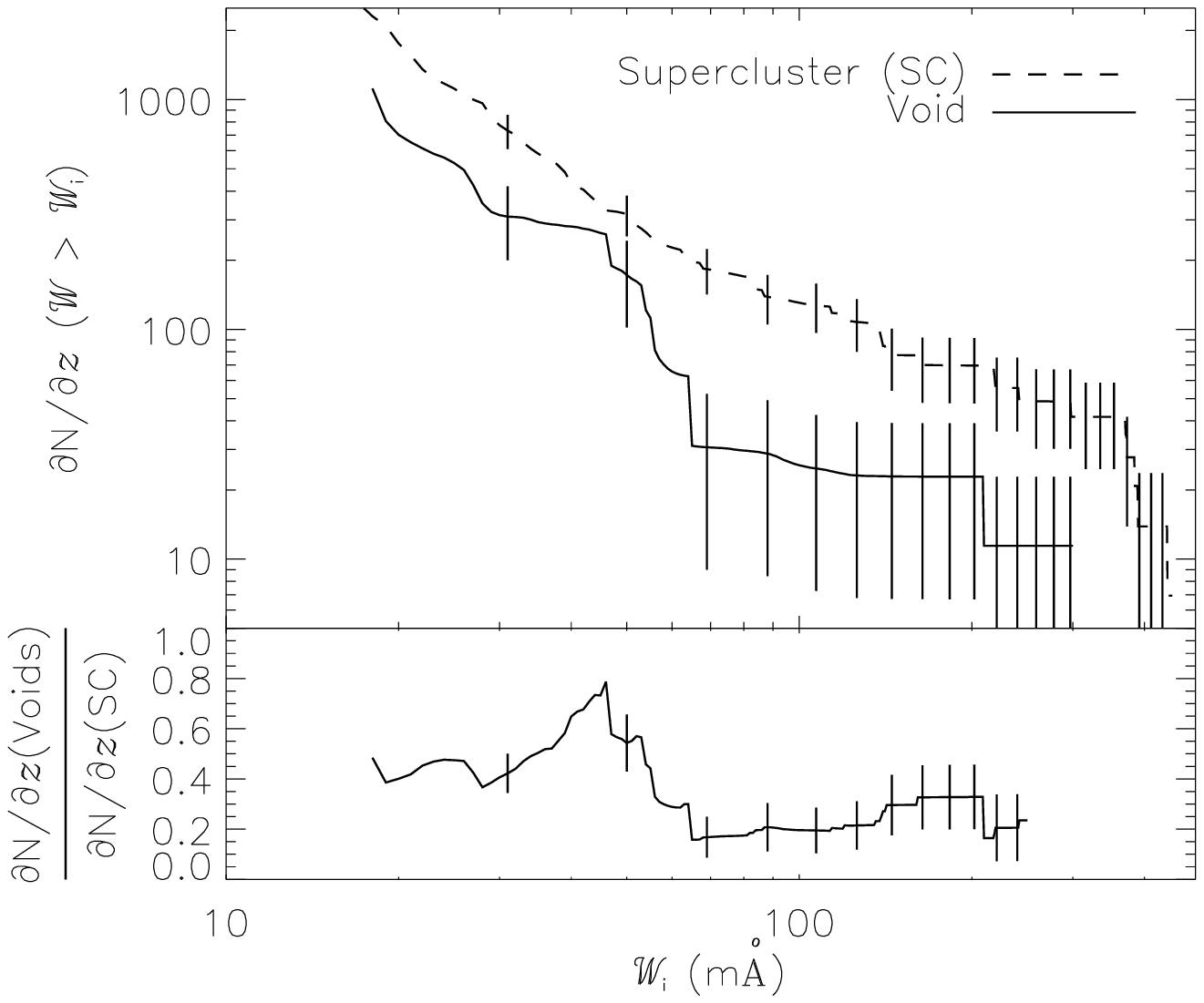}
  \caption{\label{DNDZvoid} The upper panel displays the 
\dndz spectrum in \W of \lya absorbers divided according to nearest-neighbor 
distance. The solid line in the upper panel gives \dndz for our \sample void absorber sample, 
using a ``void'' nearest-neighbor limit of 3\hsfi Mpc. 
Absorbers and regions of space not in voids are labeled as
supercluster (SC), and \dndz for these absorbers/locations are indicated by the dashed line.
Error bars (\onesigno) are displayed for every 20 bins of 1\Mang width.
The bottom panel shows \dndzno(voids) divided by \dndzno(SC). 
This ratio suggests that the ``void absorbers'' may be of systematically lower \W than the
``supercluster absorbers". However, there are only 8 ``void'' absorbers in our sample, 
an insufficient number to make any statistically significant assertions.}
\end{figure}

As in Paper~II, we take \dndz($\Wno \gt \Wi$) to be the integrated line-of-sight frequency,
\begin{equation} \dndz (\W >\Wi) = \int_{\Wi}^\infty  \dtwondWdzover~d\W .\end{equation}
The upper panel of Figure~\ref{DNDZvoid} shows \dndz($\Wno \gt \Wi$) for our \sample void absorber sample
as the solid line, while the supercluster \sample sample is shown by the dashed line.
The \Wi\ bins in this figure are 1\Mang in width, and \onesig error bars are given every 20 \Wi\ bins.
Given the current limited sample, the ratio of \dndz in superclusters to voids is consistent with a constant value
 of $3.8\pm1.3$ for all \Wno. In other words, $22 \pm 8$\% of the \lowz \lya population are ``void''
absorbers. At this time, there is no statistically significant evidence that void
and supercluster absorbers have  different column density distributions, given that there are only
8 ``void'' absorbers in our sample.
However, as can be seen in the bottom panel of Figure~\ref{DNDZvoid}, which displays
\dndzno(void) divided by \dndzno(SC=supercluster), there may be a slight 
overabundance of low \W absorbers in ``voids'. 
This is consistent with our earlier assertions (\S~\ref{sec:CDF}) that
the low \W absorbers showed a more random distribution.

We now follow the methodology given in Paper~II for determining the cosmic baryon density, \omegab. 
In Paper~II we estimated that $\sim20\%$ of the local baryonic matter may reside in these \lya absorption systems. 
The relative number density of \lya absorbers in voids  and in superclusters found herein 
allows a first determination of the baryon density in voids. 
In Paper~II, we pointed out that for isothermal spherical \lya clouds of characteristic radius, $R_{cl}$,
\begin{equation}
\omegab \propto \phi_{cl} R_{cl}^3 \propto~~(\dndzno)~R_{cl},
\end{equation}
where $\phi_{cl}$ is the space density of the \lya absorbers.
While this estimate is dependent upon both the geometry of the
absorbers and the intensity of the ionizing radiation impinging on these clouds
(see Paper II and \citet{pks} for details), these factors enter here only if they
are different for ``void absorbers'' compared to ''supercluster absorbers''.
Thus, by assuming these factors are the same in voids and superclusters,
the ratio of baryon density in clusters to that in voids is given by,
\begin{equation}
{\omegab({\rm void}) \over \omegab({\rm total})} = {\dndzno({\rm void})\over \dndzno({\rm total})}~~\left\langle {R_{\rm void}\over R_{\rm total}} \right\rangle = 4.5\pm1.5\%~\left\langle {R_{\rm void}\over R_{\rm total}} \right\rangle.
\end{equation}
Recent galaxy-void surveys have failed to turn up
significant number of galaxies in voids down to 
impressively low luminosity limits \citep[e.g., M$_B \le -13;$][]{McLin02,Hibbard02}.
Therefore, since \lya clouds are the only detectable baryons in voids, 
\baryonfraction\ of the total baryonic density in the local Universe,
inferred upon primordial nucleosynthesis, is contained in voids. 
\clearpage
\section{Two Point Correlation Functions of Galaxy Halos and \lya Absorbers}\label{sec:GGbeam}
\begin{figure}[htbp]
\epsscale{0.7}
  \plotone{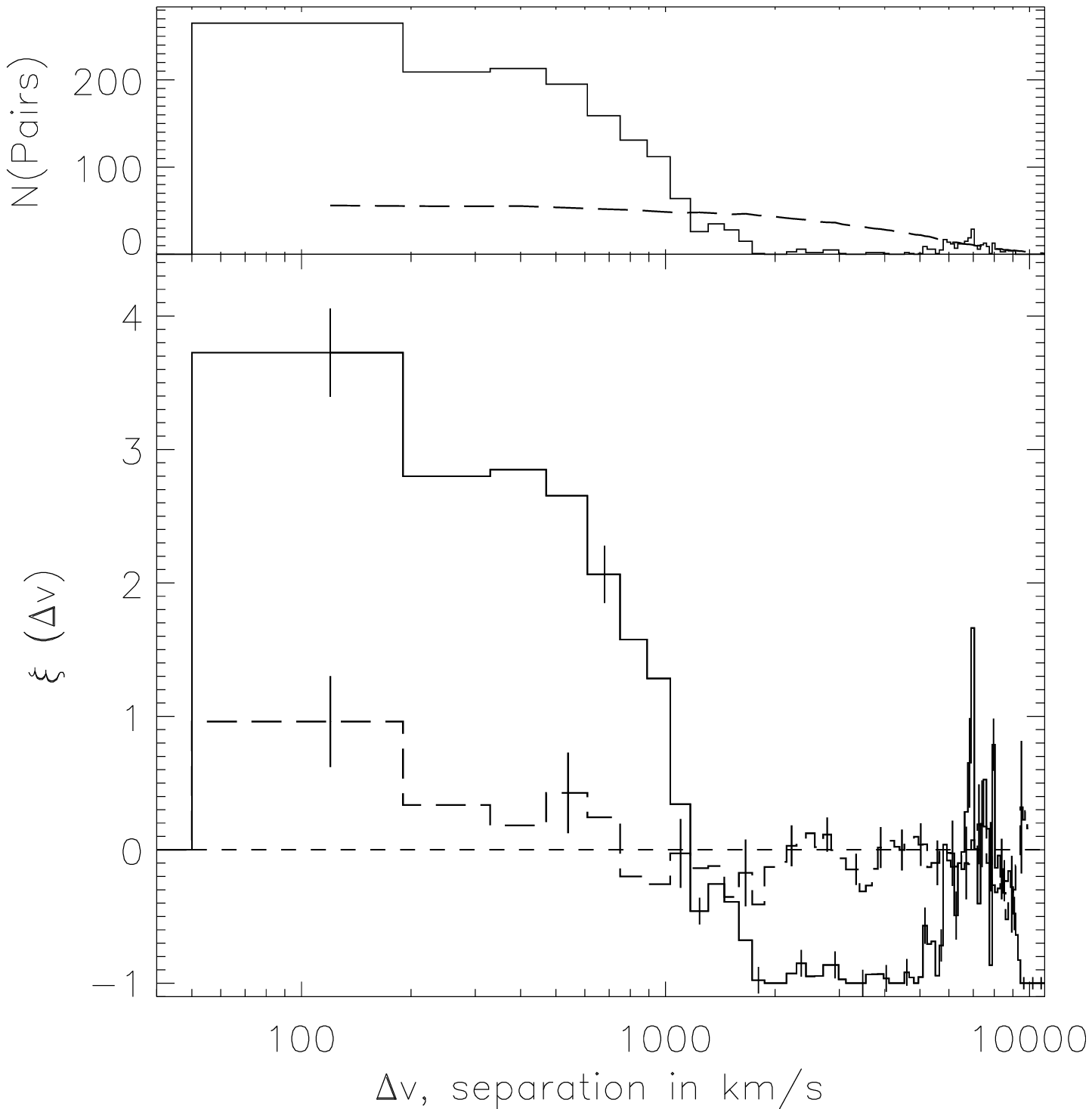}
  \caption{\label{TPCFgal} Galaxy-galaxy TPCF along our \sample 
sightlines with $\Delta v = 140$\kms bins, including only galaxies 
within 750\hsfi kpc of our HST sightlines. The upper panel compares the number, N(\Dv), 
of galaxy pairs (solid histogram) to that expected for a random 
distribution (dashed line). The bottom panel compares the TPCF, 
$\xi(\Dv)$, for galaxy halo pairs (solid histogram) to the TPCF for \lya absorber
pairs (dashed histogram).  For clarity, we plot every fifth error 
bar and indicate the random TPCF ($\xi$=0) as a horizontal short-dashed line.}
\end{figure}

In this section, we determine the characteristic width of 
\lowz galaxy filaments and voids, and compare the galaxy-galaxy  
two-point correlation function (TPCF) to the \lya cloud TPCF determined 
in Paper~II. Along each sightline, we extracted a cylindrical 
``core sample'' through the CfA survey volume of galaxies with radii of 
0.5, 0.75, and 1 \hsfi Mpc.  
The 1D TPCF of the galaxies within these cylinders was determined
using  the location along the sightline of closest approach to each galaxy. 
Therefore, this is a measure of the TPCF of extended gaseous halos along these
sightlines. As in the previous 
sections, to ensure approximately uniform galaxy information, we restrict 
our sightlines to our \sample sample, and to be consistent with 
our previous TPCFs, we do not include separations $\le$ 50\kmsno.
To obtain information about larger separations,
we do not restrict our pathlength to the available GHRS spectral coverage.
We use the full ``core sample''  from a minimum \cz =575\kms to (\proxno) from the 
target, provided that the sightline is still in a region of the CfA survey which is complete 
down to at least \Lstarno.  As in Paper~II, we normalize by the expected random 
distribution to obtain the TPCF,
$\xi(\Dv) = N_{obs} (\Dv) /  N_{ran}(\Dv) -1$.
For all trials, we see strong galaxy clustering below 2000\kmsno, 
peaking at 10\sig in the lowest separations. The TPCF for the 
750\hsfi kpc radius trial is displayed in Figure~\ref{TPCFgal} and is quite 
similar to the other trials. 

As is obvious in Figure~\ref{TPCFgal}, 
the galaxy-galaxy TPCF differs significantly from the \lya TPCF and indicates that 
\lya absorbers  cluster nowhere near as strongly as galaxy halos
would cluster along these sightlines.
Within \about 750\kmsno, the galaxy-galaxy TPCF has dropped 
to half of its peak value at the lowest separations. This leads us to 
the conclusion that the galaxy filaments in our sample are of order 750\kms in velocity 
width, or \about 11\hsfi Mpc.  Peculiar velocities (PVs) would broaden the measured galaxy TPCF, so 
this estimate of \lowz filament width is an upper limit.
If we assume a PV of $\pm$ 200\kms for galaxies in these filaments, then
the actual physical width is \about 5\hsfi Mpc.
Therefore, the filaments are either rather broad, or the peculiar 
velocities within the filaments are large.  
This result is consistent with the filament widths found in \S~\ref{sec:stat_filaments}.

While the secondary peak at $\about8000\kms$ in the galaxy TPCF is exactly 
the signal predicted by \citet{Lapp91} associated with galaxy voids 
of order 25-50 \hsfi Mpc, we caution that the TPCF presented has insufficient pathlength per
sightline to draw firm conclusions  at $\Dv \gt 3000\kmsno$.
The presence of \lya clouds in voids is evidenced in Figure~\ref{TPCFgal} by the rather
uniform velocity distribution of \lya absorbers compared to the deficiency in the galaxy-galaxy TPCF 
for $\Dv \ge 1500\kmsno$.  
Thus, the \lya absorbers and galaxy halos cluster differently in two respects:
(1) the \lya absorbers do not exhibit the large clustering signature seen for the galaxy 
halos at $\Delta v \lt 1000$\kms and 
(2) the  \lya absorbers do not exhibit the TPCF deficiency beginning at $\Delta v \sim 1500$\kms that
occurs in the galaxy TPCF.
Because the \lya absorbers and galaxy halo TPCFs are significantly different 
(the K-S probability that they are drawn from the same parent population is $< 10^{-10}$),
this gives additional support to the contention that these \lya absorbers are
not very extended galaxy halos. However, we caution that there has been a  
critique of the \lya TPCF methodology by \citet{FS96} that claims that unresolved line
blending in \lya absorbers can hide significant clustering. Since higher resolution spectra
(e.g., STIS echelle) show only modest line blending, and combined HST/FUSE Lyman line
curve-of-growth analysis requires only 2--3 components per line to explain discrepancies in
\bvalues \citep{Shull00}, we suspect that effects of blending on the TPCF are small. Further,
we see no way that this blending  can  produce a large TPCF excess at
$\Dv = 500-1000$\kmsno, 10 to 20 times the  velocity resolution of our data.
\clearpage
\section{Conclusions}\label{sec:conclusions}
	This paper is the third in a series whose purposes are to discover
very \lowzno, low column density (\Nh$\leq10^{15}$\percmtwono) \lya absorbers, 
determine their basic properties (e.g., line density, \bvalue distribution, TPCF
amplitude, etc.), and determine their relationship  with galaxies,
voids, and large-scale structure in the local Universe. The papers in this
series use \lya detections made with the GHRS aboard the HST and utilized the
G160M medium resolution first-order grating to obtain $\sim19\kmsno$ spectral resolution and limiting
\Wnos\ as low as $\sim12\mang$. Future papers in this series will
continue this investigation using newer spectra obtained with the 
Space Telescope Imaging Spectrograph (STIS) plus medium resolution gratings, which 
have similar spectral resolution and limiting \W to the GHRS data.

In this paper we have used the Revised CfA 
Galaxy Redshift Catalog 
to compare the detected \lya absorber locations to the actual galaxy 
distribution at low \z. For the full sample, distances between \lya 
absorbers and the nearest known CfA catalog 
galaxy neighbors range from $\sim100\hsfi $kpc to $\gt10\hsfi $Mpc.
Because galaxy redshift information is non-uniform across the sky,
we constructed a sub-sample of \Nsample\ $SL \ge \foursig$ \lya absorbers.
These absorbers are located in regions of the CfA catalog that  have been 
completely surveyed down to at least \Lstar galaxies to achieve some level of consistency in the 
\lyano-galaxy correlation analysis. 
For the majority of our studies, we restrict ourselves to this \sample sample. 

From the specific statistical results summarized below, we find no strong evidence that low
column density, \lowz \lya absorbers arise in very  extended galaxy halos. These absorbers may
be a different population than some or all of the higher column density absorbers studied by
L95 and C98. However, the absorbers do show a significant correlation with the large-scale
distribution of galaxies. This conclusion is consistent with numerical simulations
\citep[e.g.,][]{Dave99} of large-scale structure formation in which the low column density
\lya absorbers are true intergalactic material, not gas recycled by galaxies into the IGM.
Observational work from several other groups  (IPF99, Morris \etl 1993; Rauch \etl 1996;
Tripp \& Savage 1998, IPF99) support this interpretation.
 The results that draw us to this conclusion are:
\begin{itemize}

\item Of our \Nsample\ \sample\ HST/GHRS \lya absorbers, \Nsuper\
 are within or near large-scale galaxy structures (filaments), while \Nvoid\ are 
 in galaxy voids. These percentages are similar, whether we
 use a nearest galaxy distance of 3\hsfi Mpc, or the distribution of \Lstar galaxies to define 
the approximate regions of supercluster and void along our sightlines. 
Accounting for pathlength and sensitivity, we find that \Voidfraction\ of \lowz \lya
absorbers are in galaxy voids.

\item While there are six absorber galaxy pairs in sufficiently close proximity ($\le 230\hsfi $kpc) to an
\Lstar galaxy to be considered galaxy halos by previous work (L95, C98), the median distance
between \lya absorbers and galaxies in our \sample sample is $\about 500\hsfi $kpc, too large
to be a plausible galaxy halo.

\item In a cumulative nearest-neighbor distance distribution, our \lya absorbers lie intermediate
between random locations in space (median distance = 3\hsfi Mpc) and locations of galaxies
(median distance to nearest galaxy = 250\hsfi Mpc). The median distance between \lya
absorbers and galaxies is \about 500\hsfi kpc, twice the median separation between bright
galaxies. This is further evidence to doubt the association of \lya clouds with individual
galaxies. The strongest absorbers ($\Wno \gt 68$\mang) have nearest-galaxy distributions more
similar to galaxies, while the weaker absorbers ($\Wno \lt 68$\mang) cluster more weakly.
But the stronger absorbers still do not cluster as strongly as galaxies cluster (\foursig difference).
Although we have used different techniques for investigating the absorber-galaxy
relationship, these results are similar to those recently reported by \citet{Tripp98a} and
IPF99.

\item We find no correlation between \lya absorber equivalent width and the sightline
impact parameter with the nearest galaxy for low-\Nh absorbers, in contrast
to the \citet{Lanzetta95} and \citet{Chen98} results at higher \Nhno.
This lack of correlation is consistent with predictions of N-body+hydrodynamical simulations 
of the local \lya forest \citep{Dave99}, in which most absorbers and galaxies occupy
the same large-scale structure filament. The L95 and C98
interpretation that \lya absorbers are extended galaxy halos cannot be extrapolated to
lower \hone column densities and larger impact parameters than $\sim250$\hsfi kpc.
Likewise, our interpretation that \lya absorbers are intergalactic in nature
cannot necessarily be extrapolated to the highest column densities and
smallest impact parameters ($\le 50\hsfi $kpc).
 
\item A detailed  examination of the Q~1230+0115/3C~273 field for common absorbers, 
 separated by only 0.9\degr\ on the sky (230--550\hsfi kpc at $cz=1000-2300 $\kmsno), 
is incompatible with a model in which  spherical galaxy halos of high covering factor 
account for the observed \lya absorbers.  A possible filamentary gas structure $\gt20\hsfi
$Mpc long can contain both galaxies and absorbers, and explains almost all the \lya absorbers
in this field. This example suggests that \lya absorber studies combined with galaxy redshift
surveys have the potential to map out local filamentary structure in both galaxies and gas
(i.e., ``cosmic tomography").

\item We have compared statistically the location of \lya absorbers relative to
galaxies in known filamentary structures and find evidence, at the 5 to 12$\sigma$
significance level, that \lya absorbers are aligned with large-scale ``filamentary''
structures in the local Universe.  This result reinforces our inference, based upon
nearest-neighbor distances that these \lya absorbers are related to large-scale structure filaments,
not individual galaxies.

\item In Paper~II we presented a two-point correlation function (TPCF) for \lya absorber locations
in the local Universe which found no excess power at $\Delta(cz) \gt 200\kms$ and a \foursig excess at $50-200\kmsno$.
Thus, local \lya absorbers cluster together only very weakly, similiar to the clustering
properties of \lya absorbers at $\z \gt 2$ \citep{Rauch92}, and much less
than galaxies cluster with each other. 
The difference between the clustering of galaxies and \lya clouds was shown
explicitly in this paper by constructing a TPCF of galaxy halos. The galaxy halo TPCF
shows excess power at $\Dv < 1000$\kmsno, as well as a deficiency starting
at $\Delta v\about1500$\kmsno, which may be a  signature of galaxy voids \citep{Lapp91}. 
The \lya absorber TPCF shows neither of these signatures.
\end{itemize}

In Paper~II we estimated that $\sim20$\% of the local baryonic matter may
reside in \lya absorption systems. Using the relative number density of \lya absorbers in voids 
and in superclusters (\dndzno(voids) is \Voidfraction\ of \dndzno(total)), 
allows a first determination of the baryon density in voids: \baryonfraction\ of the total baryonic 
density inferred from primordial nucleosynthesis.
This number should provide useful constraints on simulation of large-scale structure
formation. 
Because these absorbers are the only detected material in voids, and because their
total mean baryonic content (in both superclusters and voids) is comparable to that in
galaxies, it is important to understand the details of their number density, physical
conditions, metal abundance, and distribution in space relative to galaxies. 

The HST/GHRS study presented in Papers~I through III of this series have
provided an important first step towards a detailed understanding of this important
universal component.
Future papers in this series will continue to refine the results
herein by presenting: 
(1) updated statistical values (e.g., a revised TPCF) for low column density clouds based upon both the present dataset and
 spectra obtained with STIS/G140M \citep{PaperIV}; 
(2) an updated analysis of the
mean galaxy density surrounding local \lya clouds similar to that 
conducted by \citet{Grogin98} but with many more absorbers \citep{Grogin02}, and  
(3) deep galaxy surveys in regions
surrounding the GHRS and STIS sightlines to determine if low luminosity and/or low surface brightness
galaxies are responsible for \lya absorptions, particularly the ``void
absorbers'' \citep{McLin02,Hibbard02}. 
\acknowledgements
The three papers in this series make up the bulk of the Ph.D. dissertation of S.V.P.
 at the University of Colorado, Boulder. 
For their assistance in obtaining the HST/GHRS data over several
cycles, we are grateful to the staff at the Space Telescope Science
Institute, particularly Ray Lucas.  We thank Mark Giroux for helpful
discussions and a critical reading of the manuscript, and the referee, Chris Impey,
for many helpful suggestions and criticisms.  
We also thank Jessica Rosenberg for her analysis of the Q1230+0115 STIS echelle spectrum.
This work was supported by HST guest observer grant GO-06593.01-95A,  
by the HST/COS project (NAS5-98043), and by the Astrophysical Theory Program
(NASA grant NAGW-766). 

The Digitized Sky Surveys were produced at the Space Telescope Science Institute under
U.S. Government grant NAG W-2166. The images of these surveys are based on photographic
data obtained using the Oschin Schmidt Telescope on Palomar Mountain and the UK Schmidt 
Telescope. The plates were processed into the present compressed digital form with the 
permission of these institutions.
%
%
\bibliographystyle{apj}

%
%
\end{document}